\documentclass[11pt,a4paper]{article}
\pdfoutput=1

\usepackage{graphicx}
\usepackage{epstopdf}
\usepackage{jcappub}
\usepackage{mathrsfs}
\usepackage{bbm}
\usepackage{bm}
\usepackage{dsfont}
\usepackage{tensor}
\usepackage{xspace}
\usepackage{aeguill}
\usepackage{wasysym}
\usepackage{microtype}
\usepackage{empheq}
\usepackage{ifthen}
\usepackage{amsmath}
\usepackage[titletoc]{appendix}
\usepackage{cleveref}
\usepackage{subcaption}

\newcommand{\ph}{\varphi}

\newcommand{\Ups}{\Upsilon}
\newcommand{\tUps}{\widetilde{\Upsilon}}
\newcommand{\twu}{\widetilde{w}_u}
\newcommand{\tU}{\widetilde{U}}
\newcommand{\utheta}{\underline{\theta}}
\newcommand{\btheta}{\bar{\theta}}

\newcommand{\etal}{\emph{et al.}\xspace}

\newcommand{\Gma}[3]{\Gamma^{#1}_{#2#3}}

\newcommand{\parfrac}[2]{\frac{\partial #1}{\partial #2}}

\newcommand{\tbf}[1]{\textbf{#1}}

\newcommand{\rref}[1]{(\ref{#1})}
\newcommand{\di}{\mathrm{d}}
\newcommand{\hnabla}{\hat{\nabla}}
\newcommand{\pam}{\partial_{\eta_-}}
\newcommand{\pap}{\partial_{\eta_+}}

\def \pa{\partial}
\newcommand{\tdev}{\mbox{\Large $\cdot$}}

\newcommand{\obs}{\text{o}}

\newcommand{\cst}{\mathrm{cst}}
\newcommand{\wl}{\mathscr{L}}
\newcommand{\ch}{\mathscr{C}}

\def\beq{\begin{equation}}
\def\eeq{\end{equation}}
\def\bea{\begin{eqnarray}}
\def\eea{\end{eqnarray}}
\def\bean{\begin{eqnarray*}}
\def\eean{\end{eqnarray*}}
\def \be {\begin{equation}}
\def \ee {\end{equation}}

\newcommand{\s}{\hat{s}}

\newcommand{\Acal}{\mathcal A}

\newcommand{\Ocal}{\mathcal O}

\newcommand{\Vcal}{\mathcal V}

\newcommand{\GLC}{\rm GLC}
\newcommand{\DLC}{\rm DLC}

\title{From GLC to double-null coordinates and illustration with static black holes}
\author[]{Fabien Nugier}
\affiliation[]{Leung Center for Cosmology and Particle Astrophysics, National Taiwan University, Taipei, Taiwan 10617}
\emailAdd{fnugier@ntu.edu.tw}
\date{\today}
\abstract{We present a system of coordinates deriving directly from the so-called Geodesic Light-Cone (GLC) coordinates and made of two null scalars intersecting on a 2-dimensional sphere parameterized by two constant angles along geodesics. These coordinates are shown to be equivalent to the well-known double-null coordinates. As GLC, they present interesting properties for cosmology and astrophysics. We discuss this latter topic for static black holes, showing simple descriptions for the metric or particles and photons trajectories. We also briefly comment on the time of flight of ultra-relativistic particles.}

\begin{document}

\maketitle
\flushbottom

\vspace {1cm}~
\setcounter{equation}{0}


\section{Introduction}
\label{SecIntro}

Physical coordinates have a long history in cosmology, from Temple's ``optical co-ordinates" derived in 1938 \cite{1938RSPSA.168..122T}, for which the initial motivation consisted in introducing ``some new systems of normal co-ordinates which are especially adapted to the discussion of problems of astronomical optics", to Saunders' ``observational coordinates" in 1968 \cite{saunders_observations_1968,saunders_observations_1969} and their revival with Maartens' work in 1980 \cite{Maartens1,Maartens2} (which led to applications in cosmography \cite{1985PhR...124..315E}), we can say that the idea of using physical coordinates directly related to observable quantities has been a source of concern for quite some time in the scientific community. Astrophysics and cosmology are indeed two fields for which our local observer position is complexifying our understanding of the physics. On the other hand, if one wants to address questions without relying on strong philosophical assumptions (such as the cosmological principle), the use of observation-adapted systems of coordinates can be a very good alternative.

The recent years have not been without efforts toward the goal of using coordinates directly adapted to what we measure. Observation-adapted schemes have been employed in simulations \cite{Bester:2013fya,Bester:2015gla} in order to apply the ``observational cosmology programme'' \cite{1985PhR...124..315E} in the restricted spherically symmetric dust universe case. Independently from observational motivations, the so-called geodesic light-cone (GLC) coordinates \cite{P1} were first introduced in the context of the averaging problem in cosmology \cite{Li:2008yj,Rasanen:2008be,Kolb:2009hn,Buchert:2011yu,Clarkson:2011zq,Buchert:2011sx}. They were nevertheless later employed to address tangible issues in cosmology, such as computing the effect of the large scale structure on the luminosity distance-redshift relation \cite{P2,P3,P4,P5,Marozzi:2014kua,Fanizza:2015swa}, number counts of galaxies \cite{DiDio:2014lka}, lensing \cite{Fanizza:2013doa,P6}, and the propagation of ultra-relativistic particles \cite{Fanizza:2015gdn}. It was also tested on toy models such as the Lema\^{i}tre-Tolman-Bondi \cite{P6} and Bianchi I spacetimes \cite{P7}.

We propose in the present paper another system of coordinates, close to the GLC coordinates but now using two null-like coordinates instead of one null and one time-like coordinates. This system, \emph{nicknamed} here as ``double light cone" (DLC) coordinates for convenience, shares the same nice properties as GLC. \emph{We also show that these coordinates are equivalent to the ``double-null'' coordinates of Brady, Droz, Israel and Morsink (1995)} (\emph{hence the nickname for DLC, in reference to both double-null and GLC}) \cite{Brady:1995na}. As our system of coordinates also carries some residual gauge freedoms, we explain how to fix them. This paper hence adds to the weight of interest for GLC coordinates by showing their compatibility with double-null coordinates. We also propose an illustration of these coordinates in the spirit of Temple's motivational sentence (i.e. for astrophysical objects), describing static black holes and trajectories around them, and comment on the propagation of ultra-relativistic particles.

The structure of this paper is as follows. In Sec. \ref{SecGLC} we recall facts about GLC coordinates and their most interesting properties. In Sec. \ref{SecDLC} we introduce the ``new" double light-cone coordinates (\emph{again renamed for convenience}), and study their properties in comparison to the GLC ones. Sec. \ref{SecDN} is devoted to showing that DLC coordinates are equivalent to the double-null coordinates and studying their gauge fixing. In Sec. \ref{SecDLCBH} we illustrate these coordinates first by describing static black holes (Schwarzschild and Reissner-Nordstr\"{o}m), and second, by deriving particles and photon trajectories around them. We finally comment on the time-of-flight difference between two ultra-relativistic particles in Sec. \ref{SecCommentURP}, draw some conclusions in Sec. \ref{SecConclusion}, and address some technical points in Apps. \ref{AppSecondDLCDerivation} to \ref{AppZforStaticBH}.


\section{Recalling Geodesic Light-Cone (GLC) coordinates}
\label{SecGLC}

We give a short introduction to GLC coordinates and present some of their basic properties, mainly for comparison with the double light-cone coordinates presented in Sec. \ref{SecDLC}.

\subsection{General definitions}
\label{SecGLC:GenDef}

The geodesic light-cone (GLC) coordinates $(\tau,w,\utheta^a)$ ($a=1,2$) \cite{P1} form a system of four coordinates centered on a fundamental (or ``geodesic") observer worldline. In details, $\tau$ is the proper time of this observer in geodetic motion and $w$ is a null coordinate setting the past light cones centered on this observer. Finally the angles $\utheta^a = (\theta,\phi)$ are parameterizing a topological 2-sphere $\Sigma(\tau,w)$ embedded into the intersection of the $\tau = \cst$ and $w = \cst$ hypersurfaces.

The line element in the GLC coordinates is given by \cite{P1,P7}:
\bea
\label{GLCds2}
\di s_{\GLC}^2 = \Ups^2 \di w^2-2\Ups \di w \di \tau+\gamma_{ab}(\di \utheta^a-U^a \di w)(\di \utheta^b-U^b \di w) ~~,
\eea
involving 6 arbitrary functions of $\tau$, $w$, and $\utheta^a$. These coordinates are hence perfectly general but contain a residual gauge freedom that can be fixed by simple conditions \cite{Fanizza:2013doa,P7} (see also Sec. \ref{subsecDN:gaugefixing}). The metric and its inverse, in GLC coordinates $(\tau,w,\utheta^a)$, can thus be written as:
\beq
\label{GLCmetric}
g^{\GLC}_{\mu\nu} =
\left(
\begin{array}{ccc}
0 & - \Ups &  \vec{0} \\
-\Ups & \Ups^2 + U^2 & -U_b \\
\vec0^{\,T}  &-U_a^T  & \gamma_{ab} \\
\end{array}
\right) ~~~~~,~~~~~
g_{\GLC}^{\mu\nu} =
\left(
\begin{array}{ccc}
-1 & -\Ups^{-1} & -U^b/\Ups \\
-\Ups^{-1} & 0 & \vec{0} \\
-(U^a)^T/\Ups & \vec{0}^{\, T} & \gamma^{ab}
\end{array}
\right) ~~,
\eeq
where we dropped the tildes on top of angles (as in Ref. \cite{P7}) and underlined them, differently from the notation usually employed in the ``GLC literature" (Refs. \cite{P1,P2,P3,P4,P5,P6,Marozzi:2014kua,Fanizza:2015swa,Fanizza:2013doa,DiDio:2014lka}). When needed, we will denote by $\bar\theta^a$ the homogeneous angles, i.e. the angles in an homogeneous spacetime.

\subsection{Interesting properties}
\label{SecGLC:IntProp}

There are several advantages of using the GLC coordinates. First they make light propagation very simple. Indeed, photons propagate with $(w,\utheta^a) = \vec{\cst}$ and we can define their covariant 4-momentum as $k_\mu = \pa_\mu w$, giving the contravariant $k^\mu = g_{\GLC}^{\mu\nu} k_\nu = g_{\GLC}^{\mu w} = -\delta^\mu_{\tau} / \Ups$. A direct consequence is that the geodesic deviation equation becomes trivial in these coordinate:
\beq
k^\nu \nabla_{\nu} k^\mu = 0 \quad \Rightarrow \quad \Gamma^\mu_{\tau\tau} = \frac{\pa_{\tau} \Ups}{\Ups} \delta^\mu_{\tau} \quad ,
\eeq
which is confirmed from a direct calculation of the Christoffel symbols \cite{Fanizza:2013doa}.

This simplicity translates into other quantities. The redshift of a source is for example given in terms of the metric function $\Ups$:
\beq
\label{redshift}
1+z_s = \frac{\Ups(w_\obs, \tau_\obs, \utheta^a)}{\Ups(w_\obs, \tau_s, \utheta^a)} \equiv \frac{\Ups_\obs}{\Ups_s} ~~,
\eeq
extending the homogeneous relation $1+z_s = a_\obs / a_s$ ($a$ the scale factor) to the inhomogeneous regime. Similarly, the angular distance to a source located on the observer's past light cone is:
\beq
\label{AngDist}
d_A = \frac{\gamma^{1/4}}{\sqrt{\sin \utheta}} \quad \mbox{with} \quad \gamma \equiv \det(\gamma_{ab}) = \frac{|\det (g_{\rm GLC})|}{\Ups^2} ~~,
\eeq
depending solely on the (source-located) $\gamma_{ab}$ part of the metric describing the geometry in $\Sigma(\tau,w)$. It assumes an homogeneous neighborhood for the observer (see Eq. \rref{dAandMu} otherwise).

Other advantages of GLC can be found by studying lensing from the viewpoint of the Jacobi formalism. In that case one starts with the geodesic deviation eq. (GDE)\,:
\beq
\nabla_\lambda^2 \xi^\mu = R^{\mu}_{~\,\alpha\beta\nu} k^\alpha k^\nu \xi^\beta \quad \mbox{with} \quad \nabla_\lambda\equiv {\rm D}/\di \lambda \equiv k^\mu \nabla_{\mu} \quad ,
\eeq
$\lambda$ an affine parameter along the photon path starting at a source $S$ and ending at an observer $O$, and $\xi^\mu$ an orthogonal displacement with respect to the rays led by $k^\mu$.
We project the GDE on the Sachs basis $\{ \s^\mu_A \}_{A=1,2}$ (two zweibeins with flat index $A=1,2$) satisfying:
\bea
\label{eq:Sachs}
& g_{\mu\nu} \s^\mu_A \s^\nu_B = \delta_{AB} ~~,~~ \s^\mu_A u_\mu = 0 ~~,~~ \s^\mu_A k_\mu = 0 ~~, \nonumber\\
& \Pi^\mu_\nu\nabla_\lambda \s^\nu_A = 0 ~~~~\text{with}~~~~\Pi^\mu_\nu=\delta^\mu_\nu-\frac{k^\mu k_\nu}{\left( u^\alpha k_\alpha \right)^2}-\frac{k^\mu u_\nu+u^\mu k_\nu}{u^\alpha k_\alpha} ~~.
\eea
with $u_\mu \equiv \partial_\mu \tau$ the peculiar velocity of the comoving fluid ($S$ and $O$ comoving too), $\Pi^\mu_\nu$ a ``screen" projector orthogonal to two 4-vectors:
\beq
\Pi^\mu_\nu \, u_\mu = 0 \quad , \quad \Pi^\mu_\nu \, n_\mu = 0 \quad \mbox{with} \quad n_\mu \equiv u_\mu+\left( u^\alpha k_\alpha \right)^{-1} k_{\mu} \quad .
\eeq

\begin{figure}[h!]
\centering
\begin{minipage}{9cm}
    \includegraphics[width=9cm]{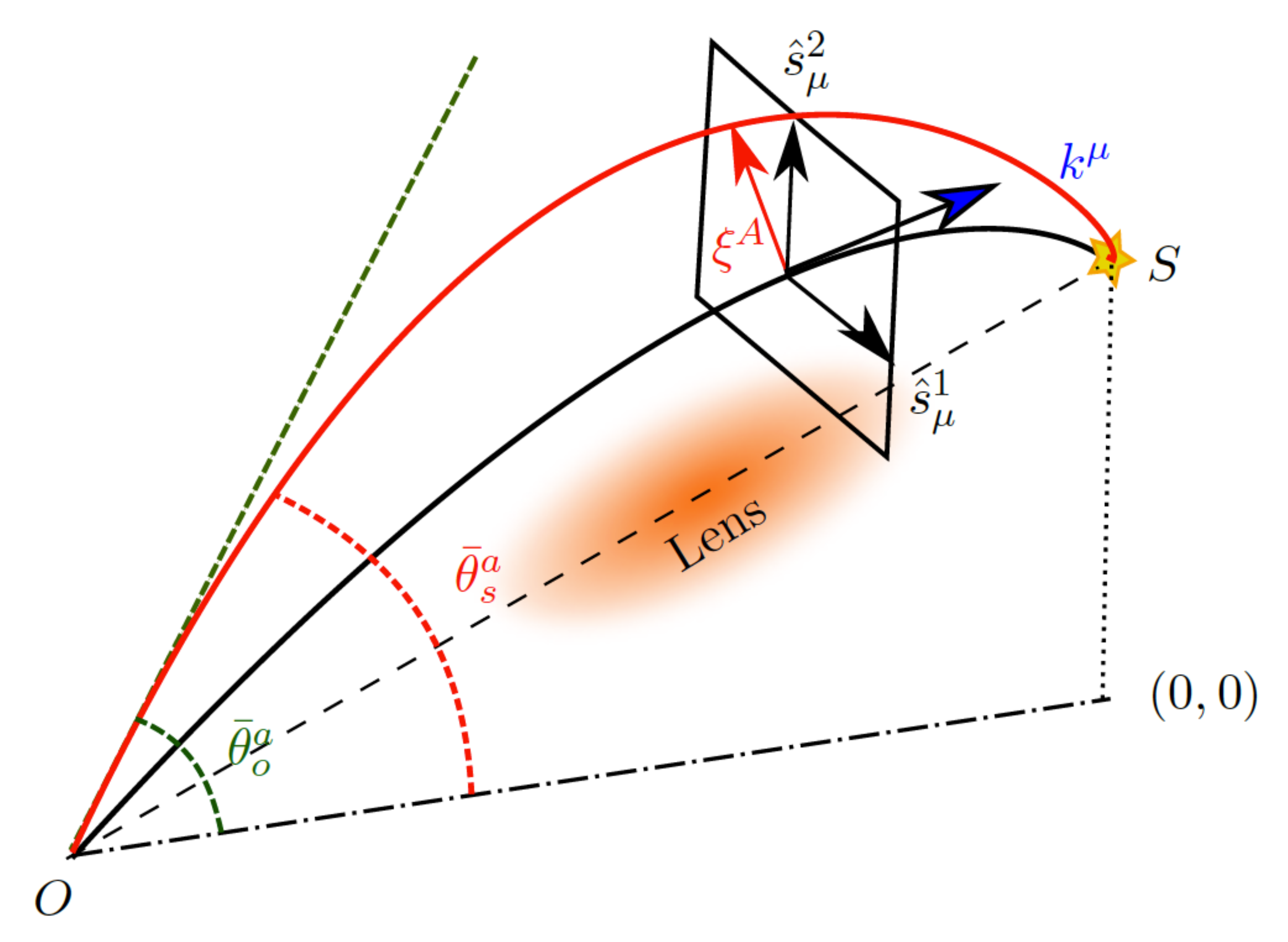}
\end{minipage}
\hfill
\begin{minipage}{5cm}
\caption{Illustration of the Jacobi map formalism in the presence of a lens. $(0,0)$ denotes the origin of angles in the sky. We present two neighbor light rays (red and black) going from $S$ to $O$. The lens is in orange.}
\label{fig:JacobiScheme}
\end{minipage}
\end{figure}  
  
We define the Jacobi map $J^A_B$ from the relation between the observed sky angle $\btheta_\obs^A$ and the screen displacement $\xi^A \equiv \xi^\mu \, \s^A_\mu$ (see Fig. \ref{fig:JacobiScheme}):
\beq
\xi^A(\lambda) = J^A_B(\lambda,\lambda_\obs) \, \btheta_\obs^A \quad .
\eeq
Projected quantities $\xi^A$ and $R^A_B \equiv R_{\alpha\beta\nu\mu}k^\alpha k^\nu \s^\beta_B \s^\mu_A$ (optical tidal matrix) bring us the Jacobi equation and its two initial conditions (see e.g. Refs. \cite{Fleury:2013sna,Fanizza:2013doa})\,:
\bea
\label{EqEvolJAB}
& \frac{\di^2}{\di \lambda^2} J^A_B (\lambda,\lambda_\obs) = R^A_C (\lambda) \, J^C_B (\lambda,\lambda_\obs) ~~, \\
\label{eq:initialConditions}
& J^A_B(\lambda_\obs,\lambda_\obs) = 0 ~~~ \mbox{ and } ~~~ \frac{\di}{\di \lambda} J^A_B (\lambda_\obs,\lambda_\obs) = (k^\mu u_\mu)_\obs \, \delta^A_B ~~.
\eea
The (unlensed or ``real'') angular position of the source $\bar\theta^A_s$ and the observed lensed position $\bar\theta^A_\obs$ (of the image) are given by\,: 
\beq
\bar\theta^A_s=\left(\xi^A / \bar d_A\right)_s \quad , \quad \bar\theta^A_\obs=\left( k^\mu\partial_\mu\xi^A / k^\mu u_\mu \right)_\obs \quad ,
\eeq
where $\bar{d}_{A}$ is the angular distance in the homogeneous and isotropic background our model refers to.
This allows us to define the so-called amplification matrix as\,:
\beq
\label{DefinitionA}
\Acal^A_B \equiv \frac{\di \btheta^A_s}{\di \btheta^B_\obs} = \frac{J^A_B (\lambda_s,\lambda_\obs)}{\bar d _A(\lambda_s)} = \left( \begin{array}{cc} 1 - \kappa - \hat{\gamma}_1 & - \hat{\gamma}_2 + \hat{\omega} \\ - \hat{\gamma}_2 - \hat{\omega} & 1 - \kappa + \hat{\gamma}_1 \end{array} \right)\\
\eeq
defining the lensing quantities $\kappa$ (convergence), $\hat\omega$ (vorticity) and $|\hat\gamma| \equiv \sqrt{(\hat\gamma_1)^2 + (\hat\gamma_2)^2}$ (shear).

In GLC coordinates, the zweibeins are written as $\hat{s}^\mu_A=(\hat{s}^\tau_A,0,\hat{s}^a_A)$ and we have $u_\mu \propto \partial_\mu \tau = \delta_{\mu}^{\tau}$ leading to $u^\mu = - \delta^\mu_\tau - \frac{1}{\Ups} \delta^\mu_w -\frac{U^a}{\Ups} \delta^\mu_a$ (for equality) and $k_\mu = \partial_\mu w$ leading to $k^\mu \equiv - \Ups^{-1}\delta^\mu_\tau$. The screen projector can thus be written as:
\beq
\Pi^\mu_\nu = \delta^\mu_\nu - \delta^\mu_\tau \delta_\nu^\tau - \delta^\mu_w \delta_\nu^w - U^a \delta^\mu_a \delta_\nu^w
\eeq
and we notice that the screen projector has no dependence from $\Ups$ or $\gamma_{ab}$.
Second, the solution to Eqs. \rref{EqEvolJAB} and \rref{eq:initialConditions} is\,:
\beq
\label{JABandCaB}
J^A_B(\lambda,\lambda_\obs) = \s_a^A(\lambda)\,\left[ 2 u_{\tau} (\dot{\gamma}_{ab})^{-1} \right]_\obs \s^B_b(\lambda_\obs)
\eeq
where $(\ldots)^{\tdev} \equiv \partial_\tau (\ldots)$\,.
The angular distance, given by
\beq
d_A(\lambda_s) = \sqrt{\det J^A_B(\lambda_s,\lambda_\obs)} \quad ,
\eeq
and the magnification $\mu \equiv 1/(\det\Acal)$, become\,:
\beq
\label{dAandMu}
d_A = \frac{2 u_{\tau_\obs} (\gamma \gamma_\obs)^{1/4}}{\sqrt{(\det \dot{\gamma}_{ab})_\obs}} \quad\quad,\quad\quad \mu = \left(\frac{\bar d_A}{d_A}\right)^2 = \frac{\Phi}{\bar{\Phi}} ~~,
\eeq
involving $\bar{d}_A$ and $\Phi$ ($\bar{\Phi}$) the flux in the in(homogeneous) geometry. 
The homogeneous distance can be chosen as $\bar{d}_A = a(\tau) r$, with $r \equiv w - \int \di \tau / a(\tau)$ measured from the observer (as in Refs. \cite{P1,P2,P3,P4,P5,P6,Marozzi:2014kua,Fanizza:2015swa,Fanizza:2013doa,DiDio:2014lka}), but that is not the only possible choice (see Ref. \cite{P7}).
Eq. \rref{dAandMu} simplifies to Eq. \rref{AngDist} when considering an homogeneous neighborhood for $O$.
\footnote{One should be careful though with the fact that both $d_A$'s numerator and denominator in Eq. \rref{dAandMu} go to zero on the observer worldline (e.g. for $r \rightarrow 0$ above). In the practical case of a perturbed FLRW geometry described by the Newtonian gauge (see App. \ref{AppDLCNearFLRW}), we find that $\gamma^{1/4} = a r (\sin\theta)^{1/2}$ and $\gamma_\obs^{1/4} / \sqrt{(\det \dot{\gamma}_{ab})_\obs} = (\sin\theta_\obs)^{-1/2}/2$. The observer angle being homogeneous ($\utheta \equiv \theta_\obs$), we get back Eq. \rref{dAandMu} at zeroth order near the observer. If first order corrections affect the observer, Eq. \rref{AngDist} is corrected with first order terms \cite{Fanizza:2013doa,Yoo:2016vne}.}
Expressions for the zweibeins can be obtained in the GLC coordinates \cite{P6}, but it is more convenient to compute the squared lensing quantities, combined with $\s^A_a \s^A_b=\gamma_{ab}$ and $\epsilon_{AB}\, \s^A_a \s^B_b=\sqrt{\gamma}\,\epsilon_{ab}$ ($\epsilon$ the anti-symmetric symbol), to get\,:
\beq
\label{LensingCombinationsInGLC}
\left( \begin{array}{c} \left( 1-\kappa \right)^2+\hat\omega^2 \\ \hat\gamma_1^2+\hat\gamma_2^2 \end{array} \right) = \left( \frac{u_{\tau_\obs}}{\bar d_A} \right)^2 \left( \left[ \frac{\gamma\,\dot\gamma_{ab}\gamma^{bc}\dot\gamma_{cd}}{\left(\det^{ab}\dot\gamma_{ab}\right)^2} \right]_\obs\gamma\,\gamma^{ad} \pm  \frac{2 \sqrt{\gamma\,\gamma_\obs}}{\left( \det^{ab}\dot\gamma_{ab} \right)_\obs} \right) ~~.
\eeq
Hence all lensing quantities are expressed with only 3 metric functions (of $\gamma_{ab}$), showing the great advantage of working in GLC coordinates.


\section{Introducing Double Light-Cone (DLC) coordinates}
\label{SecDLC}

Let us consider an observer and his/her worldline $\wl_\obs$ in a 4-dimensional Minkowski spacetime. At any given time, this observer can define a past light cone by the use of one null coordinate $w_v$ and a future light cone with another null coordinates $w_u$. There are several choices that the observer can do to define these values locally, a possible one is his/her proper time (e.g. $w_v = w_u = \tau_\obs$ \cite{P7}) or a function of it. If one considers two surfaces $w_v = \cst$ and $w_u = \cst$ such as $T_u$ (the tip of the $w_u = \cst$ cone) is in the past of $T_v$ (the tip of the $w_v = \cst$ cone) and along $\wl_\obs$, we then have an intersection of the two cones that we can denote as $\Sigma(w_u,w_v)$, a topological sphere on which we can define two angular coordinates $\theta^a$ ($a=1,2$). This is true unless the null (past and future) cones have some caustics, which is not considered here.

\subsection{Metric form}
\label{SecDLC:Metric}

Let us temporarily call $x^\mu \equiv (\tau,w,\utheta^a)$ the GLC coordinates and call $y^\mu \equiv (w_u,w_v,\theta^a)$ the new system of coordinates that we wish to satisfy the above-mentioned properties. Hence we shall now refer to these coordinates as \emph{double light-cone} (DLC) coordinates. The general transformation of coordinates between them is given by the relation:
\beq
\label{eq:TransfoCoordinates}
g_{\mu\nu}^{\DLC}(y) = \frac{\pa x^\alpha}{\pa y^\mu} \frac{\pa x^\beta}{\pa y^\nu} g_{\alpha\beta}^{\GLC}(x) \quad .
\eeq
We choose to impose $w = w_v$ as we want the DLC past light cone to match with the GLC one\footnote{Actually the choice $w_v = w$ is also a convenient choice avoiding unnecessary complications. One could for example take a modified GLC system of coordinates, spanned by $\tau$ and future light cones $w = \cst$, and then identify $w_u$ with $w$. We choose to stay as close as possible to GLC in our definition of DLC coordinates.}, so $\partial w / \partial w_v = 1$. We also want the new coordinate $w_u$ to be independent from $w_v$ and thus require that $\partial w / \partial w_u = 0$. Because in GLC we have $\utheta^a$ independent from $w$, we also impose $\partial \utheta^a / \partial w_v = 0$. This being said, one finds that the DLC metric has the following components:
\beq
\left\{
\begin{array}{ccl}
\label{TransformationGLCtoDLC}
g^{\DLC}_{w_u w_u} &=& \gamma_{ab} \parfrac{\utheta^a}{w_u} \parfrac{\utheta^b}{w_u} \quad, \\
g^{\DLC}_{w_v w_v} &=& (\Ups^2 + U^2) - 2 \Ups \frac{\pa \tau}{\pa w_v} \quad, \\
g^{\DLC}_{w_u w_v} &=& - \Ups \frac{\pa \tau}{\pa w_u} - U_a \parfrac{\utheta^a}{w_u} \quad, \\
g^{\DLC}_{w_u a} &=& \gamma_{bc} \parfrac{\utheta^b}{w_u} \parfrac{\utheta^c}{\theta^a} \quad, \\
g^{\DLC}_{w_v a} &=& -\Ups \left( \parfrac{\tau}{w_v} \parfrac{w}{\theta^a} + \parfrac{\tau}{\theta^a} \right) + (\Ups^2 + U^2) \parfrac{w}{\theta^a} - U_b \parfrac{\utheta^b}{\theta^a} \quad, \\
g^{\DLC}_{ab} &=& -2 \Ups \left( \parfrac{\tau}{\theta^a} \parfrac{w}{\theta^b} + \parfrac{w}{\theta^a} \parfrac{\tau}{\theta^b} \right) + (\Ups^2 + U^2) \parfrac{w}{\theta^a} \parfrac{w}{\theta^b} \\
& & - 2 U_c \left( \parfrac{w}{\theta^a} \parfrac{\utheta^c}{\theta^b} + \parfrac{\utheta^c}{\theta^a} \parfrac{w}{\theta^b} \right) + \gamma_{cd} \parfrac{\utheta^c}{\theta^a} \parfrac{\utheta^d}{\theta^b} \quad .
\end{array}
\right.
\eeq
We can further ask that light rays are independent from the future light-cone coordinate $w_u$. This translates into $\partial \utheta^a / \partial w_u = 0$ and thus gives:
\beq
\label{DLCmetricTMP}
g^{\DLC}_{\mu\nu} =
\left(
\begin{array}{ccc}
0 & g^{\DLC}_{w_u w_v} &  \vec{0} \\
g^{\DLC}_{w_u w_v} & g^{\DLC}_{w_v w_v} & g^{\DLC}_{w_v a} \\
\vec0^{\,T}  & (g^{\DLC}_{w_v a})^T  & g^{\DLC}_{ab} \\
\end{array}
\right)
\quad\quad \mbox{where} \quad\quad
\left\{
\begin{array}{ccc}
g^{\DLC}_{w_u w_v} = - \Ups \frac{\pa \tau}{\pa w_u} ~~, \\
g^{\DLC}_{w_v w_v} = (\Ups^2 + U^2) - 2 \Ups \frac{\pa \tau}{\pa w_v} ~~,
\end{array}
\right.
\eeq
and the angular components $g^{\DLC}_{w_v a}$ and $g^{\DLC}_{ab}$ are unchanged with respect to Eq. \rref{TransformationGLCtoDLC}.
Imposing now that the angles in DLC are equal to the ones of GLC (as it is allowed by the residual gauge freedom on $\Sigma(w_u,w_v)$, see Sec. \ref{subsecDN:gaugefixing}), we have $\partial \utheta^a / \partial \theta^b \equiv \delta^a_b$ and we further impose that $\partial w / \partial \theta^a = 0$ to get:
\beq
g^{\DLC}_{ab} = \gamma_{ab} \quad,\quad g^{\DLC}_{w_v a} = - U_a - \Ups \parfrac{\tau}{\theta^a} \equiv - \tU_a \quad.
\eeq
Taking the inverse of $g^{\DLC}_{\mu\nu}$ we obtain:
\beq
\label{DLCinvmetricTMP}
g_{\DLC}^{\mu\nu} =
\left(
\begin{array}{ccc}
g_{\DLC}^{w_u w_u} & -2/\tUps^2 & -2\tU^b/\tUps^2 \\
-2/\tUps^2 & 0 & \vec{0} \\
-2(\tU^a)^T/\tUps^2 & \vec{0}^{\, T} & \gamma^{ab}
\end{array}
\right) ~~,
\eeq
where we have introduced $\tU^a$ and $\tUps$ such that:
\beq
\label{eq:tUtUps}
\tU^a \equiv \gamma^{ab} \tU_b = U^a + \Ups \gamma^{ab} \parfrac{\tau}{\theta^b} \quad , \quad \tUps = \sqrt{2 \Ups \parfrac{\tau}{w_u}} \quad.
\eeq
We can also verify that:
\beq
\label{eq:gUwuUwu}
g_{\DLC}^{w_u w_u} = \frac{4}{\tUps^4} \left[ 2 \Ups U^a \parfrac{\tau}{\theta^a} + \Ups^2 \gamma^{ab} \parfrac{\tau}{\theta^a} \parfrac{\tau}{\theta^b} + 2 \Ups \parfrac{\tau}{w_v} - \Ups^2 \right]
\eeq
and we can see from Eq. \rref{DLCinvmetricTMP} that the only condition to make $w_u$ null is given by $g_{\DLC}^{w_u w_u} = 0$. This condition and the definition of $\tUps$ are equivalent to the following conditions on $\tau$:
\beq
\label{AnsatzTau}
\frac{\pa \tau}{\pa w_u} = \frac{\tUps^2}{2 \Ups} \quad , \quad \frac{\pa \tau}{\pa w_v} = \frac{\Ups}{2} - \tU^a \parfrac{\tau}{\theta^a} + \frac{\Ups}{2} \gamma^{ab} \parfrac{\tau}{\theta^a} \parfrac{\tau}{\theta^b} \quad .
\eeq
As we can see these conditions are not trivial and they define $\tUps$ and $\tU^a$ in a particular way. Once they are satisfied, we get the inverse metric:
\beq
\label{DLCinvmetric}
g_{\DLC}^{\mu\nu} =
\left(
\begin{array}{ccc}
0 & -2/\tUps^2 & -2\tU^b/\tUps^2 \\
-2/\tUps^2 & 0 & \vec{0} \\
-2(\tU^a)^T/\tUps^2 & \vec{0}^{\, T} & \gamma^{ab}
\end{array}
\right) ~~,
\eeq
and the direct metric is:
\beq
\label{DLCmetric}
g^{\DLC}_{\mu\nu} =
\left(
\begin{array}{ccc}
0 & - \tUps^2/2 &  \vec{0} \\
-\tUps^2/2 & \tU^2 & -\tU_b \\
\vec0^{\,T}  &-\tU_a^T  & \gamma_{ab} \\
\end{array}
\right) ~~,
\eeq
where we can appreciate the separation of $\tUps^2$ and $\tU^2$ in comparison with GLC. It is also important to notice that $\Ups$ and $U^a$ disappeared from the metric, being replaced only by $\tilde \Ups$ and $\tU^a$.
The line element in DLC coordinates $(w_u,w_v,\theta^a)$ has the following form:
\bea
\label{DLCds2}
\di s_{\DLC}^2 = -\tUps^2 \di w_u \di w_v + \gamma_{ab}(\di \theta^a-\tU^a \di w_v)(\di \theta^b-\tU^b \di w_v) ~~,
\eea
where we can notice that $\di \theta^a$ and $\tUps$ (as well as $\Ups$) are dimensionless quantities, $\di w_u$ and $\di w_v$ have dimension of a distance (assuming the speed of light $c =1$), while $\gamma_{ab}$ has the dimension of a squared distance and $U^a$ an inverse distance.

\begin{figure}[h!]
\centering
\begin{subfigure}[p]{0.45\linewidth}
    \includegraphics[width=7cm]{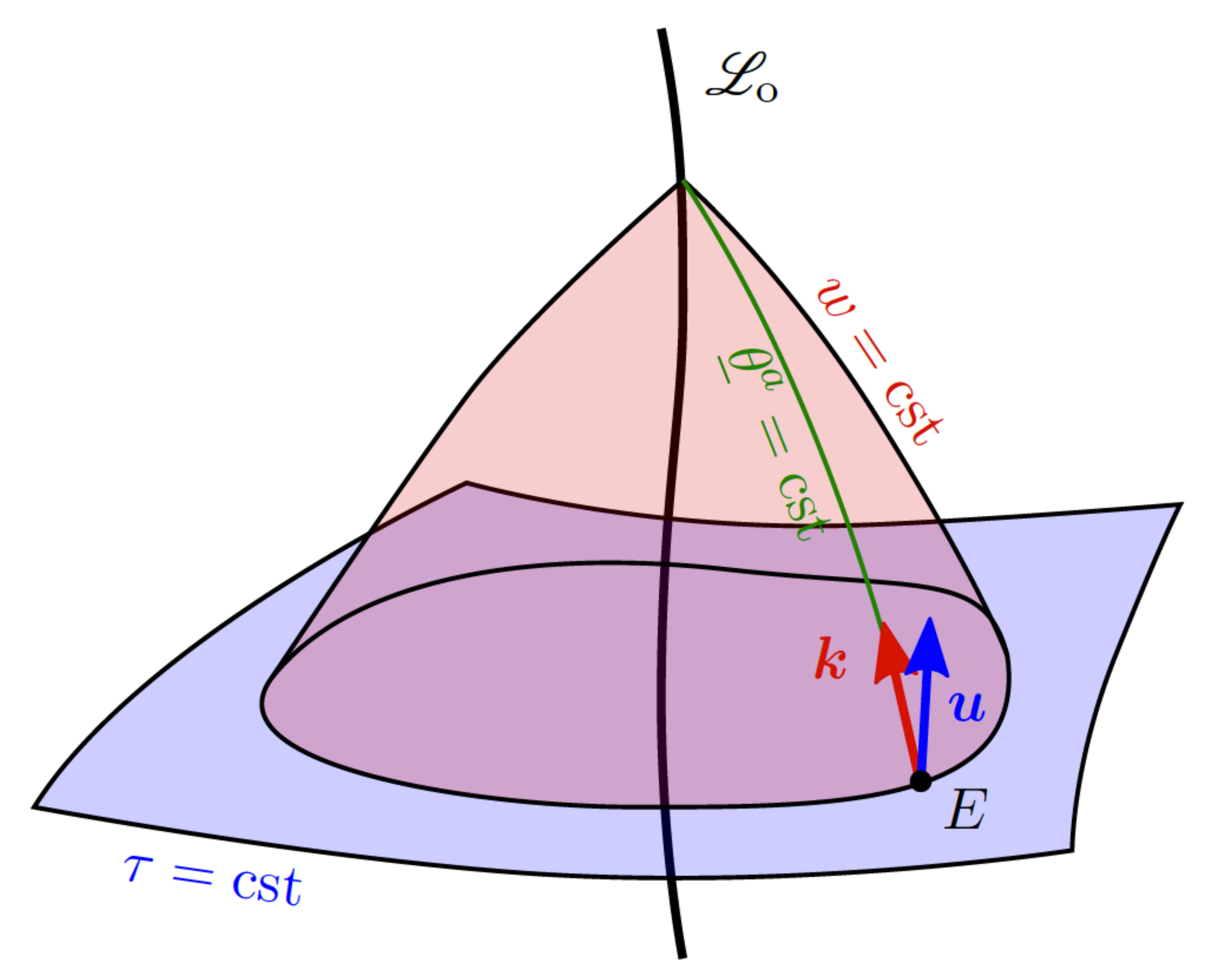}
\caption{GLC coordinates $\tau,w,\utheta^a$. The curve $\wl_\obs$ is the observer's worldline. $\tbf{k}$ represents the photon momentum and $\tbf{u}$ the peculiar velocity of a source at event position $E$.}
\label{fig:GLC_coordinates}
\end{subfigure}
\hfill
\begin{subfigure}[p]{0.45\linewidth}
    \includegraphics[width=7cm]{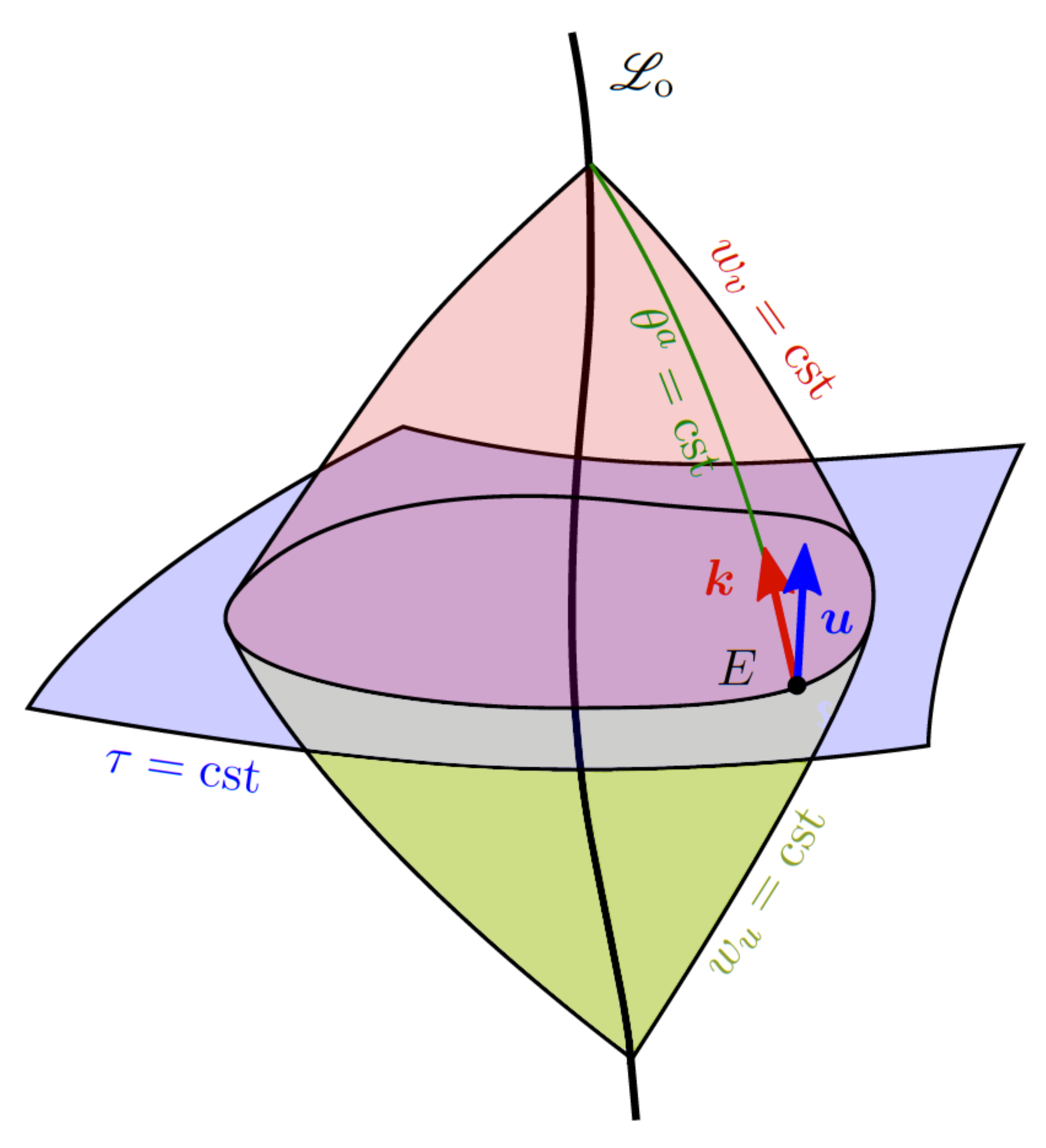}
\caption{DLC coordinates $w_u,w_v,\theta^a$.}
\label{fig:DLC_coordinates}
\end{subfigure}
\\[0.5cm]
\caption{\label{fig:GLCDLC_coordinates} Illustration of GLC (left) and DLC (right) coordinates.}
\end{figure}

To summarize, we have computed here the DLC metric from the GLC one, introducing simplifying relations along the way until a double null coordinate formulation was reached. A different derivation, based on the transformation of coordinates, is also possible. We show this derivation in App. \ref{AppSecondDLCDerivation1} and find that the two approaches are equivalent. More importantly, we can show that $w_u$ has a well defined expression in terms of GLC coordinates, and thus that GLC and DLC coordinates are perfectly consistent with each other. This derivation, made order by order in a perturbed FLRW geometry, is slightly technical and hence reported in App. \ref{AppSecondDLCDerivation2}. We also sketch the perturbative transformation of coordinates in the Newtonian gauge in App. \ref{AppDLCNearFLRW}.
Finally, we found here that the DLC coordinates replace the geodesic-observer proper time $\tau$ of GLC coordinates (see Fig. \ref{fig:GLC_coordinates}) by a null coordinate $w_u$, having for consequence to redefine the functions $\Ups$ into $\tUps$ and $U^a$ into $\tU^a$. As for the other quantities -- $\gamma_{ab}$, $w \equiv w_v$ and $\theta^a$ -- they keep the same exact definitions between the two sets of coordinates. Finally, the $w_v = \cst$ and $w_u = \cst$ hypersurfaces respectively correspond to the past and future light cones intersecting on the 2-sphere $\Sigma(w_u,w_v)$, as illustrated in Fig. \ref{fig:DLC_coordinates}.

\subsection{Simple quantities}
\label{SecDLC:SimpQuant}

We can now derive simple physical quantities directly from these new coordinates, in order to make use of them later. The photon momentum 4-vector, for example, is defined as:
\beq
\label{DLCphotonMomentum}
k_\mu \equiv \partial_\mu w_v = \delta_{\mu}^{w_v} \quad \quad \Rightarrow \quad \quad k^\mu = - \frac{2}{\tUps^2} \delta^\mu_{w_u} ~~,
\eeq
while the observer velocity, defined as in GLC coordinates $u_\mu \equiv \partial_\mu \tau$ and using Eqs. \rref{eq:tUtUps} and \rref{AnsatzTau}, is found to be:
\beq
\label{eq:DLCumu1}
u_\mu \equiv \frac{\tUps^2}{2 \Ups} \delta_{\mu}^{w_u} + \left( \frac{- \tU^2 + (\Ups^2 + U^2)}{2 \Ups} \right) \delta_{\mu}^{w_v} + \left( \frac{\tU_a - U_a}{\Ups} \right) \delta^a_\mu \quad .
\eeq
This implies that:
\beq
\label{eq:DLCumu2}
u^\mu = - \frac{\Ups}{\tUps^2} \left[ 1 + \frac{(\tU_a - U_a)(\tU^a - U^a)}{\Ups^2} \right] \delta^\mu_{w_u} - \frac{1}{\Ups}\delta^\mu_{w_v}-\frac{U^a}{\Ups} \delta^\mu_a \quad ,
\eeq
where we can see that the components $u^{w_v}$ and $u^a$ are identical to GLC. It is interesting to notice that because the observer peculiar velocity is here defined from the GLC coordinates condition $u_\mu \equiv \partial_\mu \tau$, its explicit form in DLC coordinates depends on both GLC and DLC functions $\Ups$, $\tUps$, $U^a$, and $\tU^a$. Also, $\tau$ having the dimension of a distance, we see that $u_{w_u}$ and $u_{w_v}$ are dimensionless while $u_a$ has the dimension of a distance.  The photon momentum and the observer 4-velocity lead to the product $k_\mu u^\mu = - 1 / \Ups$ and the redshift expression:
\beq
1+z_s \equiv \frac{(k_\mu u^\mu)_s}{(k_\mu u^\mu)_\obs} = \frac{\Ups_\obs}{\Ups_s}
\eeq
where $\obs$ and $s$ denote an observer $O$ (i.e. not redshifted) and a source $S$ belonging to the same past null ray. One can notice that $\tUps$ has disappeared from $k_\mu u^\mu$, hence the result, to give an expression identical to the one in GLC.

\subsection{Extra physical relations}
\label{SecDLC:ExtraRel}

From the last subsection we can see that the null geodesic equation is non-trivial only for $\mu = w_u$ and gives:
\beq
\label{eq:NGEinDLC}
k^\nu \nabla_\nu k^\mu \equiv k^\nu \pa_\nu k^\mu + \Gamma^\mu_{\alpha\beta} k^\alpha k^\beta = 0
\quad\quad \Rightarrow \quad\quad
\Gamma^\mu_{w_u w_u} = 2 \frac{\pa_{w_u} \tUps}{\tUps} \delta^\mu_{w_u} ~~.
\eeq
This is an interesting relation that we can check by a direct computation of the Christoffel symbols, as presented in App. \ref{AppChristoffel}.
On the other hand, in GLC we have $\tau$ which stands as the proper time of the observer defining a geodesic flow. We can conserve this property by imposing some conditions between the GLC and DLC metric functions. Indeed, the geodesic flow is defined by $g_{\GLC}^{\tau\tau} = -1$, thus:
\beq
u^\nu \nabla_\nu u_\mu = 0 \quad,
\eeq
and using Eqs. \rref{eq:DLCumu1} and \rref{eq:DLCumu2} we find the following evolution equations to be satisfied:
\beq
\hnabla \tUps = \frac{\tUps}{2 \Ups} \hnabla \Ups \quad , \quad \hnabla \tU_a = \hnabla U_a \quad , \quad \hnabla (\tU^2 - U^2) = \left[ \tU^2 - U^2 + \Ups (2 - \Ups) \right] \hnabla \Ups \quad ,
\eeq
with:
\beq
\hnabla \equiv \left( \frac{\Ups}{\tUps} \right)^2 \left[ 1 + \frac{(\tU_a - U_a)(\tU^a - U^a)}{\Ups^2} \right] \nabla_{w_u} + \nabla_{w_v} + U^a \nabla_a \quad .
\eeq
Finally, if we require the null energy condition to be satisfied by Einstein equations \cite{Parikh:2015wae}, we have:
\beq
\label{eq:nullenergycondition}
T_{\mu\nu} k^\mu k^\nu \geq 0 \quad \quad \Leftrightarrow \quad \quad R_{\mu\nu} k^\mu k^\nu \geq 0 \quad \quad \Leftrightarrow \quad \quad R_{w_u w_u} \geq 0 ~~.
\eeq
The component $R_{w_u w_u}$, expressed in DLC coordinates, is shown in Appendix \ref{AppChristoffel}.

\subsection{Sachs vectors}
\label{SecDLC:SachsVec}

With a view on lensing, one can introduce the Sachs basis defined in Eq. \rref{eq:Sachs} and show that the explicit expression of the screen projector $\Pi^\mu_\nu$ in DLC coordinates is:
\beq
\Pi^\mu_\nu = \delta^\mu_\nu - \delta^\mu_{w_u} \delta_\nu^{w_u} - \delta^\mu_{w_v} \delta_\nu^{w_v} + 2 \frac{(\tU^a - U^a)U_a}{\tUps^2} \delta^\mu_{w_u} \delta_{\nu}^{w_v} - 2 \frac{\tU_a - U_a}{\tUps^2} \delta^\mu_{w_u} \delta_{\nu}^a
- U^a \delta^\mu_a \delta_{\nu}^{w_v} ~~.
\eeq
It is interesting to notice that in DLC coordinates the screen projector relies mostly on its angular part and the metric functions $U^a$ and $\tU^a$. It also has a very simple expression when $\tU^a = U^a = 0$, as it is for a spherically symmetric case. We can check explicitly from Eqs. \rref{DLCphotonMomentum}, \rref{eq:DLCumu1} and \rref{eq:DLCumu2} that:
\beq
\Pi^\mu_\nu ~~\perp~~ k_\mu ~\mbox{and}~ k^\nu
\quad\quad , \quad\quad
\Pi^\mu_\nu ~~\perp~~ u_\mu ~\mbox{and}~ u^\nu \quad ,
\eeq
or any of their combinations. This is an interesting property revealing that the screen projector is orthogonal to the photon momentum and the geodesic observer peculiar velocity, as expected for such a quantity.

Writing down the conditions of Eq. \rref{eq:Sachs}, we find the relations satisfied by the Sachs vectors:
\beq
\s^{w_u}_A = - 2 \left( \frac{\tU^a - U^a}{\tUps^2} \right) \hat{s}^a_A
\quad,\quad
\s^{w_v}_A = 0
\quad,\quad
\gamma_{ab} \s^a_A \s^b_B = \delta_{AB}
\quad,\quad
\nabla_\lambda \s^{a}_A = 0 \quad ,
\eeq
with $\lambda$ and affine parameter along the photon light ray. Because $\s^\mu_A$ are defined orthogonal to $k_\mu$, they define a screen for the future light rays that the observer can emit. And as $\hat{s}^a_A$ are constant over the propagation, for which $\lambda = w_v$ is also a possible choice, we have that the evolution of $\s^{w_u}_A$ is only determined by $(\tU^a - U^a) / \tUps^2$. On the other hand, the covariant Sachs vector $\s^A_\mu = g_{\mu\nu}^{\DLC} \s^\nu_A$ is orthogonal to $k^\mu$ and thus defines a screen for past light rays received by the observer, for which we can choose $\lambda = w_u$ (like in Fig. \ref{fig:JacobiScheme}). We have the components:
\beq
\s_{w_u}^A = 0
\quad,\quad
\s_{w_v}^A = - \frac{\tUps^2}{2} \s^{w_u}_A - \tU_a \s^a_A
\quad,\quad
\s_{a}^A = \gamma_{ab} \, \s^b_A
\quad .
\eeq
Let us notice finally that for $\tU^a = U^a$ we get that the Sachs vectors $\s^{\mu}_A$ are only expressed in terms of their angular components and are hence constant between the different spheres embedded in the past and future light cones. This is also true for $\s_{\mu}^A$ when the extra condition $\tU^a = 0$ is imposed (as it is for a spherically symmetric geometry). These properties indicate that DLC coordinates may be better adapted than GLC for some specific physical applications.

\subsection{Lensing quantities}
\label{SecDLC:LensQuant}

No significant changes happen for lensing quantities when we use the DLC coordinates. Indeed, the Jacobi map formalism leading to their expression does not depend on a particular system of coordinates \cite{P6}. On the other hand, the Jacobi map of Eq. \rref{EqEvolJAB} does depend on an affine parameter $\lambda$. This affine parameter can be chosen in different ways\footnote{Note that Refs. \cite{P1,P2,P3,P4,P5} are taking $\lambda = -\tau$ while Ref. \cite{P6} is using $\lambda = \tau$.}, but one can show that $\lambda = \alpha \tau + \beta$ with $\alpha \neq 0$. Hence we obtain the lensing quantities following the same procedure as before, using the definition of the amplification matrix given in Eq. \rref{DefinitionA} with the Jacobi map that did not change (still given by Eq. \rref{JABandCaB}), and we get exactly like in GLC that the lensing quantities are given by Eq. \rref{LensingCombinationsInGLC}. Nevertheless, we should recall that $\gamma_{ab} = \hat{s}_a^A \hat{s}_b^A$ and the zweibeins take a different form in DLC coordinates with respect to GLC, so calculations may be simpler in some specific cases if we employ DLC coordinates.


\section{Double-null coordinates and gauge fixing}
\label{SecDN}

Here we compare the DLC coordinates with the well-known double-null coordinates of Ref. \cite{Brady:1995na}, describing the (2+2)-splitting of a 4-dimensional spacetime in terms of two null-like hypersurfaces and two spacelike surfaces at their intersections. In the DLC case, we have the two null hypersurfaces corresponding respectively to the past and future light cones centered on the observer worldline, denoted by $w^A = (w_u,w_v)$. We then shortly address the extra gauge fixing conditions that can be imposed to the DLC coordinates.

\subsection{DLC coordinates are double-null coordinates}
\label{subsecDN:Israel}

According to Ref. \cite{Brady:1995na} we can define generators $\ell^{(A)}$ ($A=0,1$) for the two null hypersurfaces $\Vcal^A$ defined by $w^A = \cst$. These 4-vectors are proportional to the gradient of $w^A$ and can be defined as\footnote{The factor $e^{\bar{\lambda}}$ is used instead of $e^\lambda$, as in Ref. \cite{Brady:1995na}, for the simple reason that $\lambda$ already denotes our affine parameter along null trajectories. Similarly, we replaced the null coordinates $u^A$ of Ref. \cite{Brady:1995na} by our $w^A$.}:
\beq
\ell_\alpha^{(A)} = e^{\bar{\lambda}} \partial_\alpha w^A \quad,
\eeq
which, associated with $g^{\alpha\beta} \partial_\alpha w^A \partial_\beta w^B = e^{-\bar{\lambda}} \eta^{AB}$, give the relation:
\beq
\ell_{(A)} \cdot \ell^{(B)} = g^{\alpha\beta} \eta_{AC} \ell_\alpha^{(C)} \ell_\beta^{(B)} = e^{\bar{\lambda}} \delta_A^B \quad ,
\eeq
with $\eta_{AB} \equiv \mbox{anti-diag}(-1,-1)$. For DLC, i.e. with $g^{\alpha\beta} = g^{\alpha\beta}_{\rm DLC}$, we easily find that:
\beq
\label{eq:barlambda}
\bar{\lambda} = \ln \left( \tUps^2 / 2 \right) \quad,
\eeq
and we can show that:
\beq
\ell_{(1)}^\alpha = \delta^\alpha_{w_u} \quad , \quad 
\ell_{(2)}^\alpha = \delta^\alpha_{w_v} + \tU^a \delta^\alpha_a \quad .
\eeq
The other two vectors tangent to any embedded spatial surface $\Sigma$ at the intersection of $\Vcal^0$ and $\Vcal^1$ can be chosen as $e^\alpha_{(a)} = \delta^\alpha_a$ ($a=2,3$). These vectors satisfy the relations $g_{ab}^{\rm DLC} \equiv \gamma_{ab} = e_{(a)} \cdot e_{(b)}$ (metric inside $\Sigma$) and $\ell^{(A)} \cdot e_{(a)} = 0 \quad \forall A=0,1 ~;~ a = 2,3$ (orthogonality with the null hypersurface generators), as expected.

In general, the foliation of the 4-dimensional spacetime is given by an embedding relation $x^\alpha = x^\alpha(w^A,\theta^a)$, here we chose the DLC embedding which is trivially $x^\alpha = (w^A,\theta^a)$. This choice breaks the manifest 4- and 2-dimensional covariance of the equations but guarantees that angles remain constant along both sets of generators $\ell^{(A)}$. With these simple quantities within our hands, we can derive the line element in the double-null coordinates $x^\alpha$ and compare it to the DLC one. Indeed, using the DLC metric and the relation:
\beq
\label{eq:dxalpha}
\di x^\alpha = \ell^\alpha_{(A)} \di w^A + (s^a_A \di w^A + \di \theta^a)e^\alpha_{(a)} \quad ,
\eeq
in which we introduced the shift vector $s^a_A$ (see Ref. \cite{Brady:1995na}), we get:
\beq
\di x^0 = \di w_u \quad , \quad \di x^1 = \di w_v \quad , \quad \di x^a = s^a_{w_u} \di w_u + (s^a_{w_v} + \tU^a) \di w_v + \di \theta^a \quad ,
\eeq
and these total derivatives can be used in $\di s^2 = g_{\alpha\beta}^{\rm DLC} \di x^\alpha \di x^\beta$ to bring the identities:
\beq
\label{eq:shiftvectcorresp}
s^a_{w_u} = 0 \quad , \quad s^a_{w_v} = - \tU^a \quad .
\eeq
This shows that the DLC metric functions $\tU^a$ can be interpreted as a shift vector in the (2+2) decomposition.

Reasoning only in the double-null coordinates system, we find from the orthonormality conditions of $\ell^{(A)}$ and $e_{(a)}$ that:
\beq
g_{\alpha\beta} = e^{-\lambda} \eta_{AB} \ell_\alpha^{(A)} \ell_\beta^{(B)} + g_{ab} e_\alpha^{(a)} e_\beta^{(b)} \quad .
\eeq
Combined with Eq. \rref{eq:dxalpha}, this leads to the line element in the double-null coordinates:
\beq
\di s^2 = g_{\alpha\beta} \di x^\alpha \di x^\beta = e^{\lambda} \eta_{AB} \di w^A \di w^B + g_{ab} (\di \theta^a + s^a_A \di w^A) (\di \theta^b + s^b_B \di w^B) \quad ,
\eeq
which directly gives the DLC line element of Eq. \rref{DLCds2} once we use Eqs. \rref{eq:barlambda}, \rref{eq:shiftvectcorresp}, and $g_{ab} = \gamma_{ab}$. \emph{This shows that the DLC coordinates correspond to a gauge fixing of the double-null coordinates}. More generally, we have proved that GLC coordinates are compatible with the well-known double-null coordinates under the simple transformation of Sec. \ref{SecDLC:Metric}.

\subsection{Gauge fixing of DLC coordinates}
\label{subsecDN:gaugefixing}

The DLC coordinates are general and gauge fixed from the six metric functions composing it. Nevertheless, some residual gauge freedoms remain. We are now going to analyse these extra gauge freedoms and explain how to fix them. In fact, the derivations presented here are very close to Sec. 2.3 of Ref. \cite{P7}, due to the fact that $(w_v,\theta^a)$ in DLC directly translate into $(w,\theta^a)$ in GLC coordinates. Hence $w_u$ plays in calculations almost the same role as $\tau$.

\medskip

\paragraph{Relabeling light cones:}

The GLC metric is invariant under the relabeling of light cones, $w_u \rightarrow w_u'(w_u)$ and $w_v \rightarrow w_v'(w_v)$, assuming the metric functions $\tUps$ and $\tU^a$ transform as:
\beq
\tUps \rightarrow \tUps' = \tUps \sqrt{\frac{\di w_u}{\di w_u'} \frac{\di w_v}{\di w_v'}} \quad\quad , \quad\quad \tU_a \rightarrow \tU_a' = \tU_a \frac{\di w_v}{\di w_v'} \quad .
\eeq
The dependence on both null coordinates for $\tUps$ is justified from the different role played by $\tUps$ with respect to $\Ups$ (whose transformation in GLC only depends on $w$), and we can understand this difference by looking at Eqs. \rref{AnsatzTau}. As in GLC we can use this gauge freedom to fix a condition on the observer, namely $\tUps(\wl_\obs) = 1$. By analogy, once this gauge fixing is done we can say that we are working in the \emph{temporal gauge} (see remark after though).

\medskip

\paragraph{Relabeling light rays:}

Light rays can also be relabeled when going from one sphere $\Sigma(w_u,w_v)$ to another $\Sigma(w_u',w_v')$. According to the choice made in defining the DLC coordinates, namely that the angular part of the metric is only related to the past light-cone coordinate $w_v$, such a relabeling is equivalent to the transformation $\theta^a \rightarrow \ph^a (w_v,\theta^a)$ and the DLC metric is invariant if $\gamma^{ab}$ and $\tU^a$ follow the transformation:
\beq
\gamma^{ab} \rightarrow \gamma'^{ab} = \gamma^{cd} \pa_c \ph^a \pa_d \ph^b \quad\quad , \quad\quad \tU^a \rightarrow \tU'^a = \tU^c \pa_c \ph^a - \pa_{w_v} \ph^a \quad .
\eeq
The check of this invariance is exactly the same as in GLC as $w_u$ does not play a role in it. We can thus use it like in GLC, imposing $\tU^a(\wl_\obs) = 0$, hence defining the \emph{photocomoving gauge}. The further requirements that $\theta^a$ are regular spherical angles at the observer and that the observer is non-rotating give the already GLC-defined \emph{non-rotating observational gauge}.

\medskip

\paragraph{Reparameterizing light rays:}

We have already derived the photon covariant momentum $k_\mu = \delta_\mu^{w_v}$ and its contravariant form $k^\mu = - (2 / \tUps^2) \delta^\mu_{w_u}$. Assuming a more general form $k_\mu = k_{w_v} (\pa_\mu w_v)$ and $k^\mu \propto \delta^\mu_{w_u}$, we can show that the geodesic equation $k^\nu \nabla_\nu k^\mu = 0$ imposes $\pa_{w_u} k_{w_v} = 0$ (exactly as $\pa_\tau k_w = 0$ in GLC). In a similar manner, keeping the same parameterization from one light ray to another leads to $\pa_a k_{w_v} = 0$. So $k_{w_v} = k_{w_v} (w_v)$ (\emph{isotropic affine parameterization}) and we can show that $k_{w_v}^\obs = - \omega_\obs \Ups_\obs$ as in GLC, with $\omega_\obs = - (u_\mu k^\mu)_\obs$ the pulsation of the photon evaluated at the observer. This exact similarity with GLC, despite $u_\mu$ being now given by Eq. \rref{eq:DLCumu1}, is related to $g^{w_v w_v}_{\DLC} = 0$. Imposing the \emph{static affine parameterization}, namely that the relation $\delta x^\mu = k^\mu \delta \lambda = - (2 k_{w_v}/\tUps^2_\obs) \delta \lambda \, \delta^\mu_{w_u}$  is independent from $w_v$ (here again $\lambda$ is the affine parameter of photon trajectories), results in the condition $\partial_{w_v} \left( k_{w_v} / \tUps^2_\obs \right) = - (\Ups_\obs / \tUps^2_\obs) \, \partial_{w_v} \omega_\obs = 0$. We thus have that $k_{w_v}$ is a pure constant that we can set to one as already used in Sec. \ref{SecDLC:SimpQuant} on DLC properties.

\medskip

\paragraph{Conformal transformations:}

Finally the DLC coordinates are also invariant under conformal transformations $g_{\mu\nu}^{\rm DLC} \rightarrow (g_{\mu\nu}^{\rm DLC})' = \Omega^{-2} g_{\mu\nu}^{\rm DLC}$, assuming the coordinates and metric functions change as:
\bea
& w_u' = w_u \quad , \quad w_v' = w_v \quad , \quad (\theta^a)' = \theta^a \quad , \\
& (\tUps)' = \Omega^{-1} \tUps \quad , \quad \gamma_{ab}' = \Omega^{-2} \gamma_{ab} \quad , \quad (\tU^a)' = \tU^a \quad .
\eea
As always, conformal transformations do not affect the photon trajectories.

\medskip

\paragraph{Remarks on the observer and gauges:}

It was shown in Ref. \cite{P1} that for a geodesic observer with peculiar velocity $\hat{n}^\mu = - \partial_\mu \tau \equiv - u^\mu$, the GLC coordinates near the observer vary as $\Delta x^\mu = \hat{n}^\mu \Delta \tau \equiv - u^\mu \Delta \tau$, where $u^\mu$ and the variations of coordinates are evaluated on the observer worldline $\wl_\obs$. Using DLC coordinates and Eq. \rref{eq:DLCumu2}, this leads to the relations:
\beq
\label{eq:Deltas}
\Delta w_u = \frac{\Ups_\obs}{\tUps_\obs^2} \left[ 1 + \frac{(\tU_a - U_a)_\obs(\tU^a - U^a)_\obs}{\Ups_\obs^2} \right] \Delta \tau \quad,\quad \Delta w_v = \frac{\Delta \tau}{\Ups_\obs} \quad,\quad \Delta \theta^a = \frac{U_\obs^a}{\Ups_\obs} \Delta \tau \quad .
\eeq
The first and second equalities are related to the relabeling of light cones and we can notice that the second and third are identical to the GLC case \cite{P1}. If we now require \emph{consistency conditions between GLC and DLC observers}, we can impose: $\tUps_\obs = \Ups_\obs$ and $\tU^a_\obs = U^a_\obs$. We then see that the temporal gauge requires $\Ups_\obs = 1$ and the photocomoving gauge imposes $U^a_\obs = 0$ (and so $\Delta \theta^a = 0$ as the observer sees isotropy locally). This also means, under these choices, that $\partial_{w_u} \tau = \partial_{w_v} \tau$ on the observer's worldline (as supported by Eqs. \rref{eq:Appdtaudwu} to \rref{eq:Appdwudw}).


\section{Static black holes in DLC coordinates}
\label{SecDLCBH}

Black holes have already been studied within double-null coordinates \cite{Eilon:2015axa}. We propose here to study static black holes with DLC coordinates in order to check their consistency, understand these coordinates better, and show that GLC coordinates can be used for astrophysical objects.

\subsection{Static black holes, simple transformation}
\label{SecDLCBH:StatBH}

As an illustrative exercise we can consider a static black hole described by the metric:
\beq
\label{GenSchBH}
\di s_{\rm stat.}^2 = - N^2 \di t^2 + N^{-2} \di r^2 + r^2 \di \Omega^2 ~~.
\eeq
One can introduce two null-like coordinates $(u,v)$ satisfying the differential relations \cite{Hwang:2011mn}:
\beq
\label{DiffrAndt}
\di r = r_{,u} \di u + r_{,v} \di v \quad \quad , \quad \quad \di t = \frac{N^2}{4} \left(-\frac{\di v}{r_{,u}} + \frac{\di u}{r_{,v}} \right) \quad.
\eeq
This leads to an equivalent formulation of the line element in terms of double null coordinates:
\beq
\label{GenSchBHnullcoord}
\di s_{\rm stat.}^2 = - N^2 \di u \di v + r(u,v)^2 \di \Omega^2 ~~.
\eeq

To be more explicit we can choose $N^2 = 1 - \frac{2 G M}{r}$ and we have a Schwarschild black hole metric. As for the two null coordinates $u$ and $v$, they are then respectively called the ingoing and outgoing Eddington-Finkelstein coordinates:
\beq
\label{TransfoCoordDLCstaticBH}
u = t - r^\ast \quad\quad , \quad\quad v = t + r^\ast \quad\quad , \quad\quad r^\ast = r + 2 G M \ln\left( \left| \frac{r}{2 G M} - 1 \right| \right) \quad ,
\eeq
and $r^\ast$ is the tortoise coordinate. It is then easy to check that Eq. \rref{GenSchBH}, with Eq. \rref{TransfoCoordDLCstaticBH}, is indeed giving Eq. \rref{GenSchBHnullcoord}. We can thus compare the form of Eq. \rref{GenSchBHnullcoord} with the DLC metric of Eq. \rref{DLCds2}, using that $\tU^a = 0$ in a spherically symmetric case and assuming the light cones to be centered on $r=0$ (the observer's worldline here is the black hole center's worldline). The identification of the diverse metric elements in then obvious, giving:
\bea
\label{DLCfunctionsStaticBH}
& w_u = u \quad , \quad w_v = v \quad , \quad \theta^a = \bar \theta^a \quad , \\
& \tUps = N_{\rm Sch.} = \sqrt{1 - \frac{2 G M}{r(u,v)}} \quad , \quad \tU^a = 0 \quad , \quad \gamma_{ab} = r^2(u,v) \delta_{ab} \quad,
\eea
with $\delta_{ab} = {\rm diag}( 1 , \sin^2 \theta )$ in spherical coordinates. Let us finally comment that the explicit expression of $r(u,v)$ requires to invert the following equality:
\beq
r + 2 G M \ln\left( \left| \frac{r}{2 G M} - 1 \right| \right) = \frac{v-u}{2} \quad .
\eeq

The case of a static Reissner-Nordstr\"{o}m (charged) black hole is not more complicated. It is simply given by another choice of $N$ which is $N^2 = 1 - \frac{2 G M}{r} + \frac{Q^2}{r^2}$. We thus have a perfect description of it within the DLC coordinates with:
\beq
\tUps = N_{Q} = \sqrt{1 - \frac{2 G M}{r(u,v)} + \frac{Q^2}{r(u,v)^2}} \quad , \quad \tU^a = 0 \quad , \quad \gamma_{ab} = r^2(u,v) \delta_{ab} \quad.
\eeq
Though we considered two particular cases here, this identification is correct for any static black hole, as proved in App. \ref{AppZforStaticBH}.

\subsection{Redshift for static black holes}
\label{SecDLCBH:RedshiftStatBH}

One can now present considerations on the observer proper time and redshift. We can show (see App. \ref{AppZforStaticBH}), that Eq. \rref{AnsatzTau} for a static black hole simplifies as:
\beq
\label{TauDerivN}
\partial_{w_u} \tau = \partial_{w_v} \tau = \frac{N}{2} \quad \mbox{and} \quad N=N(r(u,v)) \quad .
\eeq
We consider the specific case of a Schwarzschild metric and study the relationship between $\tau$ and $t$. We have $w_u = u = t - r^\ast$ and $w_v = v = t + r^\ast$ which, combined with Eq. \rref{TauDerivN}, lead to:
\beq
\pa_t \tau = N_{\rm Sch.} \quad, \quad \pa_{r} \tau = 0 \quad \Rightarrow \quad \tau = N_{\rm Sch.} t = \sqrt{1 - \frac{2 G M}{r}} t \quad ,
\eeq
and we used here that $r$ and $t$ are two independent coordinates. This relation is well-known in the literature and it relates the coordinate time $t$ to the proper time $\tau$ of a static observer in geodetic motion. Hence $\tau$ is again defining a geodesic flow, like in GLC. We can now give a look at the redshift in Schwarschild geometry and use $k^\mu u_\mu = - 1 / \Ups \equiv - \omega$ (photon pulsation). We then directly get the well known relation:
\beq
\label{eq:RedshitStatic1}
1+z_s = \frac{\omega_s}{\omega_\obs} = \frac{(N_{\rm Sch.})_\obs}{(N_{\rm Sch.})_s} = \sqrt{\frac{1 - \frac{2 G M}{r_\obs}}{1 - \frac{2 G M}{r_s}}} \quad .
\eeq
We can guess easily that this relation holds for any type of static black holes according to the relations $1 + z_s = \Ups_\obs / \Ups_s$ and $\Ups = N$. We prove that it is indeed true for any static black hole in App. \ref{AppZforStaticBH}. Hence for the charged black hole we can also write:
\beq
\label{eq:RedshitStatic2}
1+z_s = \frac{\omega_s}{\omega_\obs} = \frac{(N_{\rm Q})_\obs}{(N_{\rm Q})_s} = \sqrt{\frac{1 - \frac{2 G M}{r_\obs} + \frac{Q^2}{r_\obs^2}}{1 - \frac{2 G M}{r_s} + \frac{Q^2}{r_s^2}}} \quad .
\eeq
This subsection has shown that black holes can be described very conveniently in the DLC coordinates. We can feel that this must still be true for the general case of rotating black holes, but the technicality of this more complicated example is left for a future work.

\subsection{Trajectories near static black holes}
\label{SecTrajBH:static}

We consider here the trajectories of massive particles and photons around static black holes. More precisely, we start with the trajectory equations in DLC coordinates but quickly go back to $(t,r,\theta,\phi)$ coordinates in order to recover their usual form of Ref. \cite{schutz1985first}. We also show for photon trajectories the consequence of the relation $k_{\mu} = \partial_{\mu} w_v$.

\subsubsection{Massive particles}
\label{SecTrajBH:staticSub1}

Let us start with a relativistic particle of mass $m$ and energy $E$. One can find the trajectory of the particle thanks to its energy conservation. We have :
\beq
p_\mu p^\mu = - m^2 \quad ,
\eeq
which after considering the DLC metric and its reduced form for static black holes (see Eq. \rref{DLCfunctionsStaticBH}) becomes the trajectory equation:
\beq
- \tUps^2 p^{w_u} p^{w_v} + \gamma_{ab} p^a p^b = - m^2 \quad .
\eeq
Using that $\gamma_{ab}  = r^2 {\rm diag}(1 , \sin^2 \theta)$ and the symmetry of the problem that allows us to take $p^\theta = 0$, $\theta = \pi / 2$ (i.e. working in the equatorial plane), we get that :
\beq
\gamma_{ab} p^a p^b = \frac{m^2 L^2}{r^2} \quad ,
\eeq
where $L$ is the particle's angular momentum defined as $L \equiv p_{\phi} / m$.  We thus have:
\beq
p^{w_u} p^{w_v} - \left(\frac{m L}{r \tUps}\right)^2 = \left(\frac{m}{\tUps}\right)^2 \quad .
\eeq
This is a very simple expression for the trajectory of a relativistic particle that we can relate to the usual one expressed in terms of $(t,r,\theta,\phi)$ coordinates of the static black hole metric. Indeed, the first two components of the particle's momentum can be written as:
\beq
p^{w_u} = m \frac{\di w_u}{\di \tilde t} \quad , \quad p^{w_v} = m \frac{\di w_v}{\di \tilde t} \quad ,
\eeq
with $\tilde t$ the proper time of the particle along the trajectory, and using the transformation of coordinates presented in Eq. \rref{TransfoCoordDLCstaticBH} we can show (for $r>2GM$) that they are equivalent to:
\beq
\label{DLCParticleMomentum}
p^{w_u} = \frac{m}{N^2} \left( E - \frac{\di r(\tilde t)}{\di \tilde t} \right) \quad , \quad p^{w_v} = \frac{m}{N^2} \left( E + \frac{\di r(\tilde t)}{\di \tilde t} \right) \quad .
\eeq
The energy $E$ is given by $E = N^2 \di t / \di \tilde t$ as $p^t \equiv m \, \di t / \di \tilde t$ and $E \equiv - p_t / m$ (i.e. $E$ is related to the $0^{\rm th}$ component of the momentum in $(t,r,\theta,\phi)$ coordinates). These results are true for a Schwarzschild black hole as well as any static black hole. It is thus possible to simplify our trajectory using $\tUps = N$ and get:
\beq
\label{PartTrajStaticBH}
\left( \frac{\di r(\tilde t)}{\di \tilde t} \right)^2 - \left( E^2 - N^2 \left(1 + \frac{L^2}{r^2}\right) \right) = 0 \quad .
\eeq
We can then analyse the particle's trajectories by studying the sign of the second term of Eq. \rref{PartTrajStaticBH}, as done in Ref. \cite{schutz1985first} with the same exact equation.

\subsubsection{Massless particles}
\label{SecTrajBH:staticSub2}

Let us now consider the case of a massless particle, like a photon. We can first consider the energy conservation given by $k_\mu k^\mu = 0$ in DLC coordinates $(w_u,w_v,\theta^a)$. This reads, according to Eq. \rref{DLCfunctionsStaticBH} in the case of a static black hole:
\beq
- \tUps^2 k^{w_u} k^{w_v} + \gamma_{ab} k^a k^b = 0 \quad ,
\eeq
and we can replace $\tUps$ by $N$. We notice now from Eq. \rref{DLCphotonMomentum} that $k^{w_u} = - 2 / N^2$ and $k^{w_v} = 0$, hence we get the relation on the angular part of the photon momentum:
\beq
\gamma_{ab} k^a k^b = 0 \quad .
\eeq
This means that photons propagate orthogonally to the surface $\Sigma(\theta,\phi)$. It also means from the expression of $k^\mu$ that $k^\theta = k^\phi = 0$, i.e. the photon trajectory is trivial in DLC coordinates (a property shared with GLC coordinates), reducing simply to the following equation:
\beq
k^{w_u} \equiv \frac{\di w_u}{\di \lambda} = - \frac{2}{\tUps^2} \quad \Rightarrow \quad \frac{\di w_u}{\di \lambda} = - \frac{2}{N(w_u,w_v)^2} \quad ,
\eeq
which a priori involves the explicit expression of $r(w_u,w_v)$ to be solved. One can, on the other hand, come back on $(t,r,\theta,\phi)$ coordinates. For that we use Eq. \rref{TransfoFinalCoordAppB} which is valid for any static black hole, remark that $E = N^2 \di t / \di \lambda$, and we get:
\bea
\label{PhotonTrajStaticEasy}
k^{w_v} : \quad & \frac{\di t}{\di \lambda} + N^{-2} \frac{\di r}{\di \lambda} = 0 \quad \Leftrightarrow \quad \frac{\di r}{\di \lambda} = - E \\
k^{w_u} : \quad & \frac{\di t}{\di \lambda} - N^{-2} \frac{\di r}{\di \lambda} = N^{-2} \left(E - \frac{\di r}{\di \lambda}\right) = - \frac{2}{N^2} \quad \Leftrightarrow \quad \frac{\di r}{\di \lambda} = 2 + E \quad .
\eea
This directly leads to $E = -1$ that we interpret as the consequence from the fact that $k^\mu$ is a 4-vector pointing to the past. This also means that $\di r / \di \lambda = 1$ and thus $\lambda$ grows when we are going away from the observer. For incoming photons we have $\lambda$ growing, $r$ decreasing to zero, and $E > 0$, as physically expected. We can finally remark that this equation of motion is purely radial and it does not capture all the possible photon trajectories. This is explained from the fact that $k^\mu$ is here defining constant angular coordinates and null trajectories observed by the observer on his/her past light cone.

Let us alleviate this assumption and consider the most general photon momentum in DLC coordinates in order to derive all the possible photon trajectories. We thus have $\tilde k^{\mu} = (\tilde k^{w_u},\tilde k^{w_v},\tilde k^{\theta},\tilde k^{\phi})$ and the condition $\tilde k_{\mu} \tilde k^{\mu} = 0$ is:
\beq
-N^2 \tilde k^{w_u} \tilde k^{w_v} + r^2 (\tilde k^{\theta})^2 + r^2 \sin^2\theta (\tilde k^{\phi})^2 = 0 \quad .
\eeq
From the symmetry of the problem we can place ourselves in the equatorial plane, taking $\tilde k^\theta = 0$ and $\theta = \pi / 2$. The equation above hence becomes:
\beq
\tilde k^{w_u} \tilde k^{w_v} - \frac{r^2}{N^2}(\tilde k^{\phi})^2 = 0 \quad .
\eeq
We also have (in analogy with Eq. \rref{DLCParticleMomentum}):
\beq
\label{eq:kmuStaticBHTraj}
\tilde k^{w_u} = \frac{1}{N^2} \left( E - \frac{\di r(\lambda)}{\di \lambda} \right) \quad , \quad \tilde k^{w_v} = \frac{1}{N^2} \left( E + \frac{\di r(\lambda)}{\di \lambda} \right) \quad ,
\eeq
where $\lambda$ is the affine parameter describing the photon trajectory and we have used that $\di t / \di \lambda =N^{-2} E$. We thus get the well known photon trajectory in the $(t,r,\theta,\phi)$ coordinates after defining the photon momentum $L$ such that $L = \tilde k_{\phi}$ (hence $r^2 \sin^2\theta \, (\tilde k^\phi)^2 = L^2 / (r^2 \sin^2 \theta)$ simplified by $\theta = \pi / 2$), reading:
\beq
\label{PhotonTrajStaticReal}
\left( \frac{\di r(\lambda)}{\di \lambda} \right)^2 + \left(- E^2 + \frac{N^2 L^2}{r^2} \right) = 0 \quad .
\eeq
This relation is valid for any static black hole and well known in the literature \cite{schutz1985first}. We can finally notice that Eq. \rref{PhotonTrajStaticReal} gives back Eq. \rref{PhotonTrajStaticEasy} after imposing $L = 0$ (radial trajectory) and noticing the opposite sign between $E$ and $\di r / \di \lambda$ (incoming trajectories for $E>0$).

\subsubsection{Comment on redshift}
\label{SecTrajBH:CommentZSBH}

Let us do an extra comment here concerning the redshift of photon trajectories. In Sec. \ref{SecDLCBH:RedshiftStatBH} we derived the expression of the redshift for the photon trajectories defining the angular coordinates of DLC, i.e. the radial trajectories. We now find for general photon trajectories (see Sec. \ref{SecTrajBH:staticSub2}) that $u_\mu \tilde k^\mu = \frac{N}{2} (\tilde k^{w_u} + \tilde k^{w_v})$ and we can use Eq. \rref{eq:kmuStaticBHTraj} to get that:
\beq
u_\mu \tilde k^\mu = \frac{E}{N} \quad\quad\Rightarrow\quad\quad 1+z_s = \frac{E / N_s}{E / N_\obs} = \frac{N_\obs}{N_s} \quad ,
\eeq
as $E$ is a constant of motion fixed at the start of the trajectory and independent from the source or the observer. We have established the validity of Eqs. \rref{eq:RedshitStatic1} and \rref{eq:RedshitStatic2} in this more general case, showing that the redshift is also independent from the angular momentum $L$.

\section{Comment on ultra-relativistic particles}
\label{SecCommentURP}

The geodesic equation was recently considered within the framework of GLC coordinates \cite{Fanizza:2015gdn} (see also Ref. \cite{Fleury:2016mul}) in order to compute the time-of-flight difference between two ultra-relativistic (UR) particles. Using DLC coordinates, we can find the mass-shell constraint:
\beq
\label{eq:massshell}
\tUps^2 \dot{w}_u \dot{w}_v + 2 U_a \dot{w}_v \dot{\theta}^a - \gamma_{ab} \dot{\theta}^a \dot{\theta}^b + \ldots = \frac{m^2}{E^2} \quad ,
\eeq
where $m$ is the mass of the UR particle and $E$ its energy measured by the observer (at the origin of the coordinates). The dot-derivative is here taken with respect to the particle's proper time $\tilde t$. The above expression assumes a hierarchy among the coordinates derivatives:
\beq
\dot{w}_u \sim 1 \quad,\quad \dot{\theta}^a \sim \gamma^{-1} \quad,\quad \dot{w}_v \sim \gamma^{-2} \quad ,
\eeq
with $\gamma$ the Lorentz factor of the UR particle, ``\ldots'' denoting terms $\sim \Ocal(\gamma^{-3})$, and both sides of Eq. \rref{eq:massshell} are of order $\gamma^{-2}$. It is clear from the hierarchy that $w_u$ and $w_v$ do not have exactly an equivalent role in DLC coordinates. We can also understand this fact from App. \ref{AppSecondDLCDerivation2} where we find $w_u = - w + 2 \eta(\tau)$ at $0^{\rm th}$ order in perturbations around FLRW while $w_v = w$. Hence $\partial w_u / \partial \tau = 2 / a(\tau)$ at this order while $\partial w_v / \partial \tau = 0$. Using Eq. \rref{eq:massshell} brings the relation:
\beq
\label{eq:wvdot}
2 \dot{w}_v = \frac{(m^2/E^2) + \gamma^{ab} J_a J_b}{(\dot{w}_u \tUps^2/2) + \tU_a \dot{\theta}^a} \simeq \frac{2}{\dot{w}_u \tUps^2} \left( \frac{m^2}{E^2} + \gamma^{ab} J_a J_b \right) \quad ,
\eeq
where $J_a \equiv \gamma_{ac} \dot{\theta}^c$ and we used that $\dot{w}_u \tUps^2 \sim \Ocal(1) \gg \tU_a \dot{\theta}^a \sim \Ocal(\gamma^{-1})$. Considering from Ref. \cite{Fanizza:2015gdn} that $\dot{\tau} = \Ups_\obs / \Ups$ (involving the rescaling of the particle's proper time $\tilde t$) and that $\dot{w}_u / \dot{\tau} = \partial w_u / \partial \tau = 2 \Ups / \tUps^2$ (see e.g. Eq. \rref{eq:DLCmetricFunctions}), we see that we can approximate $\dot{w}_u \sim 2 \Ups_\obs / \tUps^2$ in the equation above. This leads to the expression:
\beq
\frac{\di w_v}{\di w_u} = \frac{\dot{w}_v}{\dot{w}_u} \simeq \frac{\tUps^2}{2 \Ups_\obs^2} \left( \frac{m^2}{E^2} + \gamma^{ab} J_a J_b \right) \quad .
\eeq
Integrating this equation now gives:
\beq
(w_v)_i - (w_v)_\obs = \int_{(w_u)_s}^{(w_u)_\obs} \frac{\tUps^2}{2 \Ups_\obs^2} \left( \frac{m^2}{E^2} + \gamma^{ab} J_a J_b \right) \di w_u \quad ,
\eeq
with i the particle index and we can neglect the $\gamma^{ab} J_a J_b$ contribution in the integral as we are integrating over the unperturbed geodesic (on which $J_a \sim 0$). Using that the time-of-flight difference between the two UR particle is $\Delta \tau = \tau_1 - \tau_2 = \Ups_\obs \left[ (w_v)_1 - (w_v)_2 \right]$, we get:
\bea
\label{eq:DeltaTau1}
\Delta \tau &=& \left( \frac{m_1^2}{2E_1^2} - \frac{m_2^2}{2E_2^2} \right) \int_{(w_u)_s}^{(w_u)_\obs} \frac{\tUps^2(w_u,w_\obs,\theta^a)}{\Ups_\obs} \, \di w_u \quad ,
\eea
with $\Ups_\obs \equiv \Ups((w_u)_\obs, (w_v)_\obs,\theta^a_\obs)$ (see App. \ref{AppSecondDLCDerivation2} for explicit limits at the observer).
We can also check in the homogeneous case (see e.g. App. \ref{AppDLCNearFLRW}) that the remaining integral simplifies to:
\beq
\int_{(w_u)_s}^{(w_u)_\obs} \frac{\tUps^2(w_u,w_\obs,\theta^a)}{\Ups_\obs} \, \di w_u = \int_{(\eta_-)_s}^{(\eta_-)_\obs} \frac{\tUps^2(\eta_-,w_\obs,\theta^a)}{\Ups_\obs(\eta_\obs,w_\obs,\theta^a_\obs)} \, \di \eta_- = \int_{\eta_s}^{\eta_\obs} \frac{a^2(\eta)}{a_\obs} \, \di \eta \quad ,
\eeq
giving back the homogeneous result:
\beq
\label{eq:DeltaTaHom}
\Delta \tau = \left( \frac{m_1^2}{2E_1^2} - \frac{m_2^2}{2E_2^2} \right) \int_{\tau_s}^{\tau_\obs} \frac{\di \tau}{1+z(\tau)} \quad .
\eeq
We can conclude this section by noticing that the DLC coordinates have given through Eq. \rref{eq:DeltaTau1} an equivalent result to the GLC one. This expression is interesting but does not bring a real simplification compared to GLC. Nevertheless, it shows that DLC coordinates are also able to describe particles which are not exactly on the light cone, as long as they are ultra relativistic particles (hence propagating close to the light cone).


\section{Conclusions}
\label{SecConclusion}

We have presented a system of coordinates that we derived directly from the geodesic light-cone (GLC) coordinates, replacing the proper time of the observer $\tau$ with a null coordinate $w_u$ while keeping the other three coordinates unchanged. We nicknamed these coordinates Double Light-Cone (DLC) coordinates as they make use of two null coordinates, share many of the advantages that GLC coordinates possess, and are mathematically equivalent to the well-known double-null coordinates of Brady \etal \cite{Brady:1995na}. They are thus adapted coordinates that can be employed in cosmology and for that reason we have attached importance to the description of their gauge fixing.

In the spirit of adapted coordinates, and recalling the initial motivation of Temple to describe astrophysical objects, we employed the DLC coordinates to the description of static black holes. We showed their usefulness, but this is not a surprise considering the multiple applications of double-null coordinates in this field. Hence our illustration was more a consistency check for DLC coordinates than a new result. We also showed that they are convenient to describe massive particle and photon trajectories, and we briefly commented on the time of flight of ultra-relativistic particles. It would be interesting to extend our analysis to rotating (Kerr) black holes and see if the DLC coordinates offer any simplification. We imposed the black hole to be at the center of coordinates in this paper, it would thus be interesting to see how the description changes when it is placed at a certain distance on our past light cone. We could also study strong lensing from this black hole \cite{Virbhadra:1999nm,Virbhadra:2008ws}, as seen from an observer at the center of coordinates, extending adapted coordinates beyond caustics.

Finally, in this paper we have considered the restricted case of an observer in geodesic motion in order to stay close to GLC. This imposed to write the peculiar velocity in terms of GLC metric functions, leading to expressions that were sometimes mixing DLC and GLC functions. This is not a restriction of DLC coordinates and we believe that they are adapted to cosmological or astrophysical studies as well as the GLC coordinates. Nevertheless, it is clear by definition that GLC coordinates are better adapted to a geodesic observer.  As for DLC, they should have the advantage in situations involving light emission and reception, and hence represent a complementary tool for GLC. As already said, they are equivalent to the double-null coordinates, up to an eventual residual gauge fixing, and they thus build the bridge between GLC and double-null coordinates. They are adapted to light propagation and can be used for black hole calculations. The DLC coordinates may also turn useful for other applications, such as black hole perturbations or even gravitational wave emissions. Adapted coordinates are useful and we should continue to develop them.


\section*{ACKNOWLEDGMENTS}

My research is supported by the Leung Center for Cosmology and Particle Astrophysics (LeCosPA) of the National Taiwan University (NTU). Any error appearing in these pages should only be attributed to my own responsibility.
I want to thank Prof. Gabriele Veneziano (CERN, Coll\`{e}ge de France) for giving me advice towards the construction of the DLC coordinates. I am also very grateful to Dr. Pierre Fleury (Univ. of Cape Town) for his comments on the draft and to Dr. Dong-Han Yeom (LeCosPA), Dr. Dong-Hoon Kim (Ewha Womans University) and Prof. Pisin Chen (LeCosPA) for our discussions regarding black holes. I am thankful to Dr. Hung-Yi Pu (ASIAA) for references on photon trajectories (see \url{https://odysseyedu.wordpress.com/} for his beautiful simulations) and to the anonymous referee who gave me the opportunity to improve the paper on points that were not explained well enough.
The initial idea of this work was initiated three years ago, at the end of my PhD, but only revived during the Second LeCosPA Symposium ``Everything About Gravity'' in December 2015. The coordinates I had derived at that time were different and not as well defined as DLC. I thank Prof. Costas Bachas (LPTENS) for having discussed the black hole application with me at this epoch.


\appendix
\begin{appendices}

\section{Direct DLC transformation and perturbed FLRW}
\label{AppSecondDLCDerivation}

We first present a direct derivation of the DLC inverse metric elements in terms of GLC coordinates and show that this approach is equivalent to Sec. \ref{SecDLC:Metric}. We then solve the condition that makes $w_u$ to be null, perturbatively and using the method of characteristics.

\subsection{General considerations}
\label{AppSecondDLCDerivation1}

As mentioned in Sec. \ref{SecDLC:Metric}, we can establish the link between GLC and DLC coordinates in another way. Indeed, taking the inverse relation of Eq. \rref{eq:TransfoCoordinates}, namely:
\beq
\label{eq:InvTransfoCoordinates}
g^{\mu\nu}_{\DLC}(y) = \frac{\pa y^\mu}{\pa x^\alpha} \frac{\pa x^\nu}{\pa x^\beta} g^{\alpha\beta}_{\GLC}(x) \quad ,
\eeq
and assuming the following identities:
\bea
w_v &=& w \quad \Rightarrow \quad \parfrac{w_v}{w} = 1 ~~,~~ \parfrac{w_v}{\tau} = \parfrac{w_v}{\utheta^a} = 0 \quad , \\
\theta^a &=& \utheta^a \quad \Rightarrow \quad \parfrac{\theta^a}{\utheta^b} = \delta^a_b ~~,~~ \parfrac{\theta^a}{\tau} = \parfrac{\theta^a}{w} = 0 \quad,
\eea
we obtain the relations:
\bea
\label{TransformationInverseGLCtoDLC}
& g_{\DLC}^{w_u w_u} = - \left( \parfrac{w_u}{\tau} \right)^2 - \frac{2}{\Ups} \parfrac{w_u}{\tau} \parfrac{w_u}{w} - 2 \frac{U^a}{\Ups} \parfrac{w_u}{\tau} \parfrac{w_u}{\utheta^a} + \gamma^{ab} \parfrac{w_u}{\utheta^a} \parfrac{w_u}{\utheta^b}
\quad, \\
& g_{\DLC}^{w_v w_v} = 0
\quad,\quad
g_{\DLC}^{w_u w_v} = - \frac{1}{\Ups} \parfrac{w_u}{\tau}
\quad, \\
& g_{\DLC}^{w_u a} = - \frac{U^a}{\Ups} \parfrac{w_u}{\tau} + \gamma^{ab} \parfrac{w_u}{\theta^b}
\quad,\quad
g_{\DLC}^{w_v a} = 0
\quad,\quad
g_{\DLC}^{ab} = \gamma^{ab} ~~.
\eea
Introducing $\tUps$ such that $g_{\DLC}^{w_u w_v} = -2/\tUps^2$ followed by $\tU^a$ such that $g_{\DLC}^{w_u a} = -2 \tU^a/\tUps^2$, we obtain the expressions of the DLC metric functions:
\beq
\label{eq:DLCmetricFunctions}
\tUps^2 = 2 \Ups \left[ \parfrac{w_u}{\tau} \right]^{-1}
\quad , \quad
\tU_a = \gamma_{ab} \tU^b = U_a - \Ups X_a
\quad , \quad
X_a \equiv \left[ \parfrac{w_u}{\utheta^a} \right] \left[ \parfrac{w_u}{\tau} \right]^{-1} \quad .
\eeq
These two relations can be employed in $g_{\DLC}^{w_u w_u}$ of Eq. \rref{TransformationInverseGLCtoDLC} to find that:
\beq
\label{eq:dwudw}
g_{\DLC}^{w_u w_u} = 0
\quad \Rightarrow \quad
\left[ \parfrac{w_u}{w} \right] \left[ \parfrac{w_u}{\tau} \right]^{-1} = \frac{\Ups}{2} \left( \gamma^{ab} X_a X_b - 1 \right) - U^a X_a
\quad ,
\eeq
as required for our coordinates to be double null. This last relation is a second order partial differential equation that gives $w_u$ in terms of GLC coordinates and metric functions once solved (see Sec. \ref{AppSecondDLCDerivation2}).

We also find that $g_{\DLC}^{w_u w_u}$ of Eq. \rref{TransformationInverseGLCtoDLC} is consistent with Eq. \rref{eq:gUwuUwu} under the condition:
\beq
\label{eq:Appdtaudwu}
\parfrac{\tau}{w_u} = - \parfrac{\tau}{w_v} \left[ \parfrac{w_u}{w} \right]^{-1} \quad .
\eeq
This relation is indeed verified after using Eq. \rref{eq:tUtUps} into Eq. \rref{eq:gUwuUwu} and imposing  $g_{\DLC}^{w_u w_u} = 0$, on the one hand:
\beq
\label{eq:Appdtaudwv}
\parfrac{\tau}{w_v} = \frac{- \tU^2 + (U^2 + \Ups^2)}{2 \Ups} \quad ,
\eeq
and combining Eqs. \rref{TransformationInverseGLCtoDLC} and \rref{eq:DLCmetricFunctions} and imposing $g_{\DLC}^{w_u w_u} = 0$, on the other hand:
\beq
\label{eq:Appdwudw}
\parfrac{w_u}{w} = \frac{\tU^2 - (U^2 + \Ups^2)}{\tUps^2} \quad .
\eeq
These three relations, with Eq. \rref{eq:tUtUps} to express $\partial_a \tau$, can be used in combination with Eq. \rref{eq:Deltas} to verify that:
\beq
\Delta \tau = \parfrac{\tau}{w_u} \Delta w_u + \parfrac{\tau}{w_v} \Delta w_v + \parfrac{\tau}{\theta^a} \Delta \theta^a \quad .
\eeq
We also show in App. \ref{AppDLCNearFLRW} that the Eq. \rref{eq:gUwuUwu} can be solved at first order in perturbations around an FLRW geometry. This section hence proved the consistency between the derivation based on coordinates transformation (from GLC to DLC) and the one based on the metric (presented in Sec. \ref{SecDLC:Metric}). We are now going to solve Eq. \rref{TransformationInverseGLCtoDLC} to prove that $w_u$ is well defined.

\subsection{Solution of $g_{\DLC}^{w_u w_u} = 0$}
\label{AppSecondDLCDerivation2}

Let us derive the expression of $w_u$ in terms of GLC coordinates and metric functions $(\tau, w, \utheta^a)$. The equation to be satisfied is given by $g_{\DLC}^{w_u w_u} = 0$ from Eq. \rref{TransformationInverseGLCtoDLC} that we simply write as:
\beq
\label{eq:dtaudwOfwu}
\partial_\tau w_u + \frac{2}{\Ups} \partial_w w_u = - 2 U^a (\Ups^{-1}) \partial_a w_u + \gamma^{ab} (\partial_a w_u)(\partial_b w_u) (\partial_\tau w_u)^{-1} \quad ,
\eeq
where $\partial_a$ denotes a derivative with respect to $\utheta^a$. This equation is a priori a non-linear partial differential equation, but an expansion of $w_u$ in perturbations around an homogeneous FLRW spacetime allows to solve it as a linear partial differential equation. Indeed, writing:
\beq
\label{eq:wuexpansion}
w_u(\tau,w,\utheta^a) = \sum_{n=0}^\infty w_u^{(n)}(\tau,w,\utheta^a) \quad ,
\eeq
we have the zeroth order $w_u^{(0)}(\tau,w,\utheta^a) = w_u^{(0)}(\tau,w)$ independently from angles (homogeneous solution). The direct consequence of that is:
\beq
\label{eq:Order0Conditions}
U^{a(0)} = 0 \quad , \quad \partial_a w_u^{(0)} = 0 \quad ,
\eeq
and the RHS of Eq. \rref{eq:dtaudwOfwu} is expressed in terms of lower orders of $w_u$ than in the LHS. In other words, Eq. \rref{eq:dtaudwOfwu} can be written at $\Ocal(n \geq 1)$ as:
\beq
\label{eq:dtaudwOfwuOrdern}
\partial_\tau w_u^{(n)} + \frac{2}{a(\tau)} \partial_w w_u^{(n)} = Y^{(n)} + Z^{(n)}
\eeq
where $Y^{(n)}$ is a contribution accounting for the difference between $\frac{2}{\Ups} \partial_w w_u$ and $\frac{2}{a} \partial_w w_u$ on the LHS and $Z^{(n)}$ is from the RHS of Eq. \rref{eq:dtaudwOfwu}:
\bea
Y^{(n)} &=& - 2 \sum_{k=1}^{n} (\Ups^{-1})^{(k)} (\partial_w w_u)^{(n-k)} \quad\quad \forall \quad n \geq 1 \quad , \\
\label{eq:Zn}
Z^{(n)} &=& - 2 \sum_{N+M+K=n} \left[ U^{a(N)} (\Ups^{-1})^{(M)} (\partial_a w_u)^{(K)} \right] \nonumber \\
& & + \sum_{N+M+K+L=n} \left[ \gamma^{ab(N)} (\partial_a w_u)^{(M)}(\partial_b w_u)^{(K)} ((\partial_\tau w_u)^{-1})^{(L)} \right]  \quad \forall \quad n \geq 2 ~~ . \quad
\eea

More precisely, we can derive a solution of Eq. \rref{eq:dtaudwOfwuOrdern} order by order. At $\Ocal(0)$ (using Eq. \rref{eq:Order0Conditions}):
\beq
\partial_\tau w_u^{(0)} + \frac{2}{a(\tau)} \partial_w w_u^{(0)} = 0 \quad .
\eeq
This is a linear partial differential equation that can be solved through the method of characteristics. We get that $w_u^{(0)}$ is constant along the characteristic curve:
\beq
\label{eq:C0}
\ch^{(0)}: \quad \quad w_\obs = -w + 2 \eta(\tau) \quad\quad \mbox{where} \quad\quad \eta(\tau) \equiv \int_{0}^{\tau} \frac{\di \tau'}{a(\tau')} \quad ,
\eeq
and its value is given in terms of a general function $\twu^{(0)}$:
\beq
w_u^{(0)} = \twu^{(0)}(w_\obs) \quad\quad \Leftrightarrow \quad\quad w_u^{(0)}(\tau,w) = \twu^{(0)}(-w+2\eta(\tau)) \quad .
\eeq
The same reasoning can be applied at $\Ocal(1)$, $\Ocal(2)$ and so one, with for example at first and second orders:
\bea
Y^{(1)} &=& \frac{2 \Ups^{(1)}}{a^2} \partial_w w_u^{(0)} \quad , \\
Z^{(1)} &=& 0 \quad , \\
Y^{(2)} &=& 2\left(\frac{\Ups^{(2)}}{a^2} - \frac{(\Ups^{(1)})^2}{a^3} \right) \partial_w w_u^{(0)} + 2\frac{\Ups^{(1)}}{a^2} \partial_w w_u^{(1)} \quad , \\
Z^{(2)} &=& - \frac{2}{a} U^{a(1)} \partial_a w_u^{(1)} + \frac{a}{2} \gamma^{ab(0)} \partial_a w_u^{(1)} \partial_b w_u^{(1)} \quad ,
\eea
where $\partial_w w_u^{(0)}$, and $\partial_w w_u^{(1)}$ or $\partial_a w_u^{(1)}$, are respectively given from the resolution of zeroth and first order equations.

At $\Ocal(n)$, the solution of Eq. \rref{eq:dtaudwOfwuOrdern} is found as follows. First we notice from the LHS that the characteristic curve is the same as the zeroth order, $\ch^{(1)} = \ch^{(0)}$. We can thus integrate along this curve and find that:
\beq
\label{eq:SolWun}
w_u^{(n)}(\tau,w,\utheta^a) = \int_{0}^{\tau} \di \tau' \left[ Y^{(n)} + Z^{(n)} \right] (\tau',-w_\obs+2\eta(\tau'),\utheta^a) + \twu^{(n)}(w_\obs) \quad ,
\eeq
where $w_\obs$ needs to be replaced by $-w+2\eta(\tau)$ and $\twu^{(n)}$ is an arbitrary function. Summing all orders and defining the functions of $(\tau,w,\utheta^a)$:
\beq
Y = \sum_{n=1}^\infty Y^{(n)} \quad , \quad Z = \sum_{n=1}^\infty Z^{(n)} \quad , \quad \twu = \sum_{n=1}^\infty \twu^{(n)} \quad ,
\eeq
we get the general solution:
\beq
\label{eq:SolWu}
w_u(\tau,w,\utheta^a) = \int_{0}^{\tau} \di \tau' \left[ Y + Z \right] (\tau',w-2\eta(\tau)+2\eta(\tau'),\utheta^a) + \twu(-w+2\eta(\tau)) \quad ,
\eeq
of the equation equivalent to Eq. \rref{eq:dtaudwOfwu}:
\beq
\label{eq:LikedtaudwOfwu}
\partial_\tau w_u + \frac{2}{a(\tau)} \partial_w w_u = Y + Z \quad .
\eeq

We now need to fix the boundary condition of $w_u$ in order to set $\twu$. In GLC we can impose the gauge condition $w |_{\wl_\obs} = \eta(\tau)$ (see e.g. Refs. \cite{Fanizza:2015swa,NugierThesis}), leading to $w_\obs |_{\wl_\obs} = \eta(\tau)$. Imposing this condition and requiring that:
\beq
w_u |_{\wl_\obs} = \eta(\tau)
\eeq
we get from Eq. \rref{eq:SolWu} that:
\beq
\label{eq:SolWu2}
\twu(x) = x - \int_{0}^{\eta^{-1}(x)} \di \tau' \left[ Y + Z \right] (\tau',-x+2\eta(\tau'),\utheta^a) \quad .
\eeq
where we used $x \equiv \eta(\tau)$ for clarity. We now have an explicit form for $\twu$ and the final expression of $w_u$ is given by:
\beq
\label{eq:SolWuExplicit}
w_u(\tau,w,\utheta^a) = - w + 2 \eta(\tau) + \int_{\tau_\obs}^{\tau} \di \tau' \left[ Y + Z \right] (\tau',w-2\eta(\tau)+2\eta(\tau'),\utheta^a) \quad ,
\eeq
where we have defined $\tau_\obs \equiv \eta^{-1}(-w+2\eta(\tau))$. We can check that $\tau |_{\wl_\obs} = \tau_\obs$, so this lower bound corresponds to the proper time of the observer on his/her own worldline. Hence the property $w_u |_{\wl_\obs} = \eta(\tau)$ is easily checked and this is also equal to $\eta(\tau_\obs)$. In another gauge we would have a different form for $\twu$ and thus $w_u$. For example the temporal gauge condition imposes $w |_{\wl_\obs} = \tau$ and we could also choose $w_u |_{\wl_\obs} = \tau$. Nevertheless in that case the expression of $\twu(x)$, with now $x \equiv \tau + \eta(\tau)$, involves the expression of $\tau(x)$ which is not easy to obtain. Hence it is better not to use the temporal gauge in that case.

Another form of Eq. \rref{eq:SolWuExplicit} solution could be obtained by integrating over $w$ rather than $\tau$. Skipping the details but noticing that the characteristic curve $\ch^{(0)}$ is unchanged, we find:
\beq
\label{eq:SolWuExplicit2}
w_u(\tau,w,\utheta^a) = - w + 2 \eta(\tau) + \frac{1}{2} \int_{w_\obs}^{w} \di w' \, a(\tau(w_\obs,w')) \left[ Y + Z \right] (\tau(w_\obs,w'),w',\utheta^a) \quad ,
\eeq
where $w_\obs \equiv - w + 2 \eta(\tau) = \eta(\tau_\obs)$ and this is consistent with the boundary conditions $w |_{\wl_\obs} = \eta(\tau)$ and $\tau |_{\wl_\obs} = \tau_\obs$ expressed above. We can check directly that $w_u |_{\wl_\obs} = \eta(\tau)$. We also defined the function $\tau(w_\obs,w') \equiv \eta^{-1}(\frac{w_\obs+w'}{2})$ for notation convenience. Let us trivially notice that the solutions of Eq. \rref{eq:SolWuExplicit} or \rref{eq:SolWuExplicit2} indeed work when plugged back into Eq. \rref{eq:LikedtaudwOfwu} (and this property is independent from the imposed boundary conditions on $\wl_\obs$). We have thus proved in this appendix that $w_u$ can be expressed in terms of GLC coordinates, at least at a perturbative level around FLRW. This, in addition to other relations presented in the paper (e.g. in Sec. \ref{SecDLC:Metric}), shows that DLC and GLC coordinates are perfectly consistent with each other. This is a non-trivial result in which we replaced the time coordinate $\tau$ by the null coordinate $w_u$ while keeping the three others identical ($w_v \equiv w$, $\theta^a \equiv \utheta^a$).

\section{DLC coordinates and the Newtonian gauge}
\label{AppDLCNearFLRW}

We show in this section some relations for the DLC coordinates and metric functions near a perturbed FLRW geometry in the Newtonian gauge. This gauge is defined with the following line element:
\beq
\di s_{\rm NG}^2 = a^2(\eta) \left[ -(1+ 2\Phi) \di \eta^2 + (1-2\Psi)(\di r^2 + r^2 \di \theta^2 + r^2 \sin^2 \theta \, \di \phi^2) \right]
\eeq
involving the so-called conformal time $\eta$ and radius $r$ (in addition to the homogeneous angles $\btheta^a = (\theta,\phi)$). The metric functions $\Phi$ and $\Psi$ are the so-called Bardeen potentials that we will later assume equal (and denote by $\psi(\eta,r,\btheta^a)$) at first order in perturbations (with no anisotropic stress, otherwise see Ref. \cite{Marozzi:2014kua}), and we neglect vectors or tensor modes (cf. e.g. \cite{P5,Fanizza:2015swa}). We can establish the transformation of coordinates between $y^\mu = (w_u ,w_v, \theta^a)$ and $x^\mu = (\eta, r, \overline{\theta}^a)$, using Eq. \rref{eq:InvTransfoCoordinates} with now $g_{\rm NG}^{\alpha\beta}$ replacing $g^{\alpha\beta}_{\GLC}$. With the first order decomposition:
\beq
w_u = \eta - r + w_u^{(1)} \quad,\quad w_v = \eta + r + w_v^{(1)} \quad,\quad \theta^a = \overline{\theta}^a + \theta^{a(1)} \quad ,
\eeq
we find the DLC metric functions at zeroth order to be:
\beq
\tUps^{(0)} = a \quad,\quad \tU^{a(0)} = 0 \quad,\quad \gamma^{ab(0)} = a^{-2} {\rm diag}(r^{-2} , r^{-2} (\sin\theta)^{-2}) \quad .
\eeq
At first order the coordinates transformations and DLC functions are:
\bea
& \pap w_u^{(1)} = \pam w_v^{(1)} = \frac{\Phi + \Psi}{2}
\quad,\quad
\pam \theta^{a(1)} = 0
\quad, \\
& \tUps^{(1)} = \frac{2}{a} \left[ \Phi - \Psi - \pam w_u^{(1)} - \pap w_v^{(1)} \right]
\quad,\quad
\tU^{a(1)} = \frac{1}{2} \left( 2 \pap \theta^{a(1)} - \gamma^{ab(0)} \partial_b w_u^{(1)} \right)
~, \\
& \gamma^{ab(1)} = 2 a^{-2} \left[ \Psi \gamma^{ab(0)} + \gamma^{ac(0)} \partial_c \theta^{b(1)} \right] \quad ,
\eea
where we have introduced the null-cone-like (but not exactly null) coordinates:
\beq
\eta_\pm = \eta \pm r \quad,\quad \partial_\eta = \pap + \pam \quad,\quad \partial_r = \pap - \pam \quad .
\eeq

We can now study the condition $g^{w_u w_u}_{\DLC} = 0$ and see if the transformations above respect it. To achieve this, one can study either Eq. \rref{eq:gUwuUwu} or \rref{TransformationInverseGLCtoDLC} perturbatively. The second relation was already studied in Sec. \ref{AppSecondDLCDerivation2}, so we consider the first approach here. Based on Eq. \rref{eq:gUwuUwu}, we define the perturbative quantities in GLC and DLC coordinates:
\bea
& \Ups = a + \Ups^{(1)} \quad , \quad U^a = U^{a(1)} \quad , \\
& \Ups^{(1)} = a(\eta) (\partial_r P - \pap Q) \quad , \quad U^{a(1)} = \partial_\eta \theta^{a(1)} - \frac{1}{a} \gamma^{ab(0)} \partial_b \tau^{(1)} \quad , \\
\label{eq:GLCOrder1}
& \tau = \tau^{(0)} + \tau^{(1)} \equiv \int_{\eta_{in}}^{\eta} \di \eta' a(\eta') + a(\eta) P(\eta,r,\btheta^a) \quad , \quad w = \eta_+ + Q \quad ,
\eea
where these results were proved in Refs. \cite{P4,NugierThesis,Fanizza:2015swa} and we define the integrals:
\small
\beq
P(\eta,r,\btheta^a) = \int_{\eta_{in}}^{\eta} \di \eta' \frac{a(\eta')}{a(\eta)} \Phi(\eta',r,\btheta^a) \quad , \quad Q(\eta_+,\eta_-,\btheta^a) = \int_{\eta_\obs}^{\eta_-} \di x \, \left( \frac{\Phi + \Psi}{2} \right) (\eta_+,x,\btheta^a) \quad .
\eeq
\normalsize
We find that Eq. \rref{eq:gUwuUwu} is trivial at zeroth order (using that $\partial_{w_v} \tau = a / 2$), as expected, and find the conditions for first and second order:
\bea
& \partial_{w_v} \tau^{(1)} = \frac{\Ups^{(1)}}{2} \quad , \\
& \partial_{w_v} \tau^{(2)} = \frac{\Ups^{(2)}}{2} - U^{a(1)} \parfrac{\tau^{(1)}}{\btheta^a} - \frac{a(\eta)}{2} \gamma^{ab(0)} \parfrac{\tau^{(1)}}{\btheta^a} \parfrac{\tau^{(1)}}{\btheta^b} \quad ,
\eea
in which we already made simplifications according to the order in perturbations. 

Let us prove that the first order relation is verified. Indeed, we can write:
\bea
\parfrac{\tau}{w_v} &=& \parfrac{\tau}{\eta_+} \parfrac{\eta_+}{w_v} + \parfrac{\tau}{\eta_-} \parfrac{\eta_-}{w_v} + \parfrac{\tau}{\btheta^a} \parfrac{\btheta^a}{w_v}\quad , \\
&=& \left( \frac{a}{2} + \parfrac{\tau^{(1)}}{\eta_+} \right) \left( 1 + \parfrac{\eta_+^{(1)}}{w_v} \right) + \frac{a}{2} \parfrac{\eta_-^{(1)}}{w_v} + \Ocal(\psi^3) \quad , \\
&=& \frac{a}{2} \left( 1 + \partial_r P + 2 \parfrac{\eta^{(1)}}{w_v} \right) \quad ,
\eea
where we used that $\partial \tau^{(1)} / \partial \eta_+ = a(\eta) \partial_r P / 2$ and $\eta_+^{(1)} + \eta_-^{(1)} = \eta^{(1)}$. Considering now that $w = \eta_+ + w^{(1)} = w_v$, we get that:
\beq
\parfrac{\eta_+^{(1)}}{w_v} = - \parfrac{w^{(1)}}{\eta_+} + \Ocal(\psi^2) = - \partial_+ Q + \Ocal(\psi^2) \quad ,
\eeq
as $w^{(1)} = Q$ from Eq. \rref{eq:GLCOrder1}. This proves that:
\beq
\parfrac{\tau^{(1)}}{w_v} = \frac{a}{2} \left( \partial_r P - \pap Q \right) \equiv \frac{\Ups^{(1)}}{2} \quad ,
\eeq
and thus Eq. \rref{eq:gUwuUwu} appears to be consistent with GLC also at first order in (scalar) perturbations around FLRW.

\section{Christoffel symbols in DLC coordinates}
\label{AppChristoffel}

In this section we present the Christoffel symbols necessary to derive Einstein equations within DLC coordinates (a goal that we do not intend to fulfill here). We use the metric and its inverse presented in Eqs. \rref{DLCinvmetric} and \rref{DLCmetric}, plus the definition of the Christoffel symbols:
\beq
\Gma{\mu}{\nu}{\rho} = \frac{g^{\mu\lambda}_{\DLC}}{2} \left( g_{\lambda\nu,\rho}^{\DLC} + g_{\lambda\rho,\nu}^{\DLC} - g_{\nu\rho,\lambda}^{\DLC} \right) \quad .
\eeq
This gives us the following components:
\bea
& \Gma{u}{u}{u} = \frac{2 \tUps_{,u}}{\tUps} \quad , \quad \Gma{v}{v}{v} = \frac{2 \tUps_{,v}}{\tUps} + \frac{(\tU^2)_{,u}}{\tUps^2} \quad , \quad \Gma{u}{u}{v} = - \frac{(\tU^a)_{,u} \tU_a}{\tUps^2} - \frac{\tU^a \tUps_{,a}}{\tUps} \quad , 
\nonumber \\
& \Gma{v}{u}{u} = 0 \quad , \quad
\Gma{u}{v}{v} = - \frac{(\tU^a)_{,v} \tU_a}{\tUps^2} +  \frac{\tU^a (\tU^2)_{,a}}{\tUps^2} \quad , \quad
\Gma{v}{u}{v} = 0 \quad , \nonumber \\
& \Gma{u}{u}{a} = \frac{\tUps_{,a}}{\tUps} + \frac{(\tU_a)_{,u}}{\tUps^2} - \frac{\tU^b (\gamma_{ab})_{,u}}{\tUps^2} \quad , \quad
\Gma{a}{u}{u} = 0 \quad , \quad
\Gma{v}{v}{a} = \frac{\tUps_{,a}}{\tUps} - \frac{(\tU_a)_{,u}}{\tUps^2} \quad , \quad \nonumber \\
& \Gma{a}{v}{v} = \frac{\tUps_{,v}}{\tUps} \tU^a + \frac{(\tU^2)_{,u}}{\tUps^2} \tU^a - \gamma^{ab} (\tU_b)_{,v} - \frac{1}{2} \gamma^{ab} (\tU^2)_{,b} \quad , \nonumber \\
& \Gma{u}{v}{a} = - \frac{(\tU^2)_{,a}}{\tUps^2} - \frac{\tU^b}{\tUps^2} \left( (\tU_a)_{,b} - (\tU_b)_{,a} + (\gamma_{ab})_{,v} \right) \quad , \quad
\Gma{v}{u}{a} = 0 \quad , \nonumber \\
& \Gma{a}{u}{v} = \frac{\gamma^{ab}}{2} \left( \tUps \tUps_{,b} - (\tU_b)_{,u} \right) \quad , \nonumber \\
& \Gma{u}{a}{b} = \frac{1}{\tUps^2} \left( (\tU_a)_{,b} + (\tU_b)_{,a} + (\gamma_{ab})_{,v} \right) - \frac{\tU^c}{\tUps^2} \left( \gamma_{ca,b} + \gamma_{cb,a} - \gamma_{ab,c} \right) \quad , \nonumber \\
& \Gma{v}{a}{b} = \frac{(\gamma_{ab})_{,u}}{\tUps^2} \quad , \quad
\Gma{a}{u}{b} = \frac{1}{2} \gamma^{ac} (\gamma_{cb})_{,u} \quad , \nonumber \\
& \Gma{a}{v}{b} = \frac{\tU^a}{\tUps^2}\left( \tUps \tUps_{,b} - (\tU_b)_{,u} \right) + \frac{1}{2} \gamma^{ac} \left( (\tU_b)_{,c} - (\tU_c)_{,b} + (\gamma_{cb})_{,v} \right) \quad , \nonumber \\
& \Gma{a}{b}{c} = \frac{\gamma^{ad}}{2} \left( \gamma_{db,c} + \gamma_{dc,b} - \gamma_{bc,d} \right) \quad ,
\eea
where, just for notational convenience, we replaced $(w_u,w_v)$ by $(u,v)$ and used the coma notation for partial derivative. We recall also that $\tU^2 \equiv \tU_a \tU^a$. The four components $\Gma{\mu}{u}{u}$, standing for $\Gma{\mu}{w_u}{w_u}$, confirm our result of Eq. \rref{eq:NGEinDLC}.

Using now the expression of the Ricci tensor:
\beq
R_{\alpha\beta} = \Gma{\rho}{\alpha}{\beta,\rho} - \Gma{\rho}{\alpha}{\rho,\beta} + \Gma{\rho}{\lambda}{\rho} \Gma{\lambda}{\beta}{\alpha} - \Gma{\rho}{\lambda}{\beta} \Gma{\lambda}{\rho}{\alpha} \quad ,
\eeq
we find that the component $R_{w_u w_u}$ is given by:
\beq
R_{w_u w_u} = \left( \frac{\tUps_{,w_u}}{\tUps} - \frac{1}{2} \right) \gamma^{ac} (\gamma_{ac})_{,w_u} -\frac{1}{4} \gamma^{bc} \gamma^{ad} (\gamma_{ac})_{,w_u} (\gamma_{db})_{,w_u} \quad .
\eeq
The null energy condition of Eq. \rref{eq:nullenergycondition} then gives a relation between the metric functions:
\beq
2  \left( \frac{2 \tUps_{,w_u}}{\tUps} - 1 \right) \gamma^{ac} (\gamma_{ac})_{,w_u} \quad \geq \quad \gamma^{bc} \gamma^{ad} (\gamma_{ac})_{,w_u} (\gamma_{db})_{,w_u} \quad .
\eeq

\section{Transformation of coordinates for static black holes}
\label{AppZforStaticBH}

We present here a general proof of the correspondence between the DLC gauge and the static black hole metric. This also gives a rather simple proof of the redshift expression $1+z_s = \frac{N_\obs}{N_s}$ for any static black hole. Let us recall that the DLC metric is given by Eq. \rref{DLCds2} while the static black hole metric is given by Eq. \rref{GenSchBHnullcoord} in terms of ingoing and outgoing null coordinates $(u,v)$. We can still assume, without loss of generality, that $w_u = u$, $w_v = v$, $\theta^a = \btheta^a$, $\tU^a = 0$ and $\gamma_{ab} = r^2(u,v) \delta_{ab}$ like presented in Eq. \rref{DLCfunctionsStaticBH}. The comparison between the two metrics is thus reduced to their ``radial'' part (as opposed to ``angular''):
\beq
\label{eq:AppMetrics}
\di s^2_{\rm DLC} = - \tUps^2 \di w_u \di w_v \quad ~~ \mbox{to compare with}\quad ~~ \di s^2_{\rm stat.} = - N^2 \di t^2 + \frac{\di r^2}{N^2} = - N^2 \di u \di v ~~.
\eeq
This clearly identifies $\tUps = N$ for static black holes, but does not give the expressions of $\Ups$. For this reason we introduce the following change of coordinates:
\beq
\label{eq:AppTransfoCoord}
\di \tau = \frac{\tUps^2}{2 \Ups} \di w_u + \frac{\Ups}{2} \di w_v \quad\quad,\quad\quad \di r = A \di w_u + B \di w_v \quad ,
\eeq
where the first relation comes from Eq. \rref{AnsatzTau} with $\partial \tau / \partial \theta^a = 0$ (due to spherical symmetry), between GLC and DLC coordinates, and the second relates the static black hole radial distance $r$ to the DLC coordinates. We further impose that the proper time of GLC coordinates is directly related to the cosmic time $t$ of the static black hole metric by $\di \tau = C \di t$. Inverting the system of Eq. \rref{eq:AppTransfoCoord} and plugging the expressions in Eq. \rref{eq:AppMetrics}, we find that:
\beq
A = -\frac{\tUps^2 N}{2 \Ups} \quad,\quad B = \frac{\Ups N}{2} \quad,\quad C = N \quad.
\eeq
Hence we already found, as expected from Sec. \ref{SecDLCBH:RedshiftStatBH}, that the proper time of the observer $\tau$ is related to the time $t$, leading to the redshift expression:
\beq
\di \tau = N \di t \quad \Rightarrow \quad \ 1+z_s = \frac{N_\obs}{N_s} \quad .
\eeq
We also established the transformation between $(t,r)$ and $(w_u,w_v)$:
\beq
\di t = \frac{\tUps^2}{2 \Ups N} \di w_u + \frac{\Ups}{2N} \di w_v \quad\quad,\quad\quad
\di r = - \frac{\tUps^2 N}{2 \Ups} \di w_u + \frac{\Ups N}{2} \di w_v \quad ,
\eeq
that we can now combine with the general transformation of Eq. \rref{DiffrAndt} (assuming again $w_u = u$, $w_v = v$). This gives:
\beq
r_{,u} = - \frac{N^3}{2\Ups} = - \frac{\tUps^2 N}{2 \Ups} \quad,\quad r_{,v} = \frac{\Ups N}{2} = \frac{\Ups N^3}{2 \tUps^2} \quad \Rightarrow \quad r_{,u} r_{,v} = - \frac{N^4}{4} \quad.
\eeq
This already confirms that $\tUps = N$ and we can impose that $r_{,u} = - r_{,v}$ to establish that:
\beq
r_{,v} = \frac{N^2}{2} = - r_{,u} \quad\quad,\quad\quad \Ups = \tUps = N \quad\quad,
\eeq
for static black holes, confirming results of Secs. \ref{SecDLCBH:StatBH} and \ref{SecDLCBH:RedshiftStatBH} and giving the useful relations:
\beq
\label{TransfoFinalCoordAppB}
\di w_u = \di t - \frac{\di r}{N^2} \quad\quad,\quad\quad \di w_v = \di t + \frac{\di r}{N^2} \quad .
\eeq

\end{appendices}


\appendix

\bibliographystyle{JHEP.bst}
\bibliography{bibliography}

\end{document}